\documentclass[11pt]{article}

\usepackage{a4wide}
\usepackage{xspace}
\usepackage{multicol}
\usepackage{amsmath}
\usepackage{graphicx}
\usepackage{epsfig}
\usepackage{hyperref}
\usepackage{url}
\usepackage{amssymb}

\newcommand{\be}{\begin{equation}}
\newcommand{\ee}{\end{equation}}

\newcommand{\AuAu}{AuAu\xspace}
\newcommand{\PbPb}{PbPb\xspace}

\newcommand{\bea}{\begin{eqnarray}}
\newcommand{\eea}{\end{eqnarray}}

\newcommand{\jet}{\mathrm{jet}}
\newcommand{\as}{\alpha_s}

\newcommand{\ie}{\emph{i.e.}\xspace}
\newcommand{\eg}{\emph{e.g.}\xspace}
\newcommand{\cnf}{\emph{cf.}\xspace}

\def\order#1{{\cal{O}}\left(#1\right)}

\newcommand{\avg}[1]{\left\langle #1 \right\rangle}

\newcommand{\hard}{\mathrm{hard}}

\newcommand{\GeV}{\,\mathrm{GeV}}
\newcommand{\TeV}{\,\mathrm{TeV}}

\def\as{\alpha_s}

\newcommand{\pthard}{\ensuremath{p_t^{pp}}}
\newcommand{\ptfull}{\ensuremath{p_t^{AA}}}
\newcommand{\pthardsub}{\ensuremath{p_t^{pp,{\rm sub}}}}
\newcommand{\ptfullsub}{\ensuremath{p_t^{AA,{\rm sub}}}}
\def\fastjet{{FastJet}\xspace}

\title{\textbf{Jet Reconstruction in Heavy Ion Collisions}}
\author{
  Matteo Cacciari,\!$^{1,2}$ Juan Rojo,$^3$ Gavin P.~Salam$^{1,4,6}$ and Gregory Soyez$^{4,5}$ \\
\\
{\sl  \small $^1$LPTHE, UPMC Univ.~Paris 6 and CNRS UMR 7589, Paris, France}\\[2pt]
{\sl  \small $^2$Universit\'e Paris Diderot, Paris, France}\\[2pt]
{\sl  \small $^3$INFN, Sezione di Milano, Milan, Italy}\\[2pt]
{\sl  \small $^4$CERN, Department of Physics, Theory Unit, CH-1211 Geneva 23, Switzerland}\\[2pt]
{\sl  \small $^5$Institut de Physique Th\'eorique, CEA Saclay, CNRS URA 2306, F-91191 Gif-sur-Yvette, France}\\[2pt]
{\sl  \small $^6$Department of Physics, Princeton University, Princeton, NJ 08544, USA}
}
\date{}

\begin{document}

\maketitle

\vspace{-9.0cm}
 \begin{flushright}
   CERN-PH-TH/2010-223\\
   October 2010\\
 \end{flushright}
\vspace{8cm}

\begin{abstract}
  We examine the problem of jet reconstruction at heavy-ion colliders
  using jet-area-based background subtraction tools as provided by
  FastJet.
  We use Monte Carlo simulations with and without quenching to study
  the performance of several jet algorithms, including the option of
  filtering, under conditions corresponding to RHIC and LHC
  collisions.
  We find that most standard algorithms perform well, though the
  anti-$k_t$ and filtered Cambridge/Aachen algorithms have clear
  advantages in terms of the reconstructed $p_t$ offset and
  dispersion.
\end{abstract}

\clearpage

\tableofcontents
\newpage

%
\section{Introduction}

Since the appearance of the first high-energy colliders in the
1980's, the study of ``jets'' of particles
produced in the final state has proved to be a powerful tool for
probing the underlying elementary dynamics of the strong force as
described by Quantum Chromodynamics (QCD).
Jets have been extensively studied at $e^+e^-$ (LEP, SLC), $ep$ (HERA)
and $p\bar p$/$pp$ (Tevatron, RHIC, LHC) colliders, with a wide
variety of jet algorithms, including infrared and collinear (IRC) safe
algorithms such as those of refs.~\cite{kt,cam,antikt,siscone}.

A topic of current interest is the use of jets in heavy-ion (HI)
collisions, where, for example, they can be used to probe the hot,
dense medium.
In the past few years, the Relativistic Heavy Ion Collider (RHIC) has
amassed significant numbers of copper-copper and gold-gold collisions
at nucleon-nucleon centre of mass energies up to
$\sqrt{s_{NN}}=200\GeV$, and the Large Hadron Collider (LHC) should
deliver high yields of lead-lead collisions at much higher
energies in the near future.

The main obstacle to studying jets in HI collisions is the presence of
the huge background given by the underlying event (UE) produced
simultaneously with the hard nucleon-nucleon collision that initiates
the high-transverse momentum jet of interest. This UE needs to be
properly subtracted from the momentum of a given jet in order to
reconstruct its ``true'' momentum, \ie the one it would have in the
absence of the UE contribution. This problem is of course well known,
and is also present, though to a much smaller extent, in jet studies
in proton-proton collisions. Various approaches to address it have
been proposed (see \eg \cite{CMS-sub,AliceHija,ATLAS08,Salur:2008hs,Lai:2008zp,Grau:2008ed,Lai:2009ai,Alice0910,Star0911}).

In ref.~\cite{subtraction} two of us proposed a jet area-related
technique to determine the transverse momentum density of a
sufficiently uniformly distributed background and to subtract it from
the jet momenta. 
The method of \cite{subtraction} introduced several novel steps, such
as the measurement of jet areas, and procedures to determine the
transverse-momentum density of the underlying event and/or pileup.
Subsequent work of ours \cite{areas,css} has sought to provide firmer
foundations for these concepts and methods, as well as practical tests
in $pp$ jet reconstruction tasks with simulated
events~\cite{optimisation}.
Preliminary experimental jet measurements from the STAR collaboration
at RHIC, whose analysis is partially based on the ideas of
\cite{subtraction}, have been presented in ref.~\cite{star}.

In this article we give a systematic examination of the performance of
such methods for heavy-ion collisions, applying them to Monte Carlo
simulations for RHIC and the LHC. In particular, we determine the
accuracy with which jet momenta can be effectively reconstructed,
comparing the performance of several different jet algorithms.

\section{Challenges of jet reconstruction in heavy-ion collisions}

Jets in HI collisions are produced in an environment that is far from
conducive to their detection and accurate measurement. 
Monte Carlo simulations (and real RHIC data) for gold-gold collisions
at $\sqrt{s_{NN}} = 200$~GeV (per nucleon-nucleon collision)
show that the transverse momentum density $\rho$ of final-state
particles is about 100 GeV per unit area (in the rapidity-azimuth
plane).
For lead-lead collisions at $\sqrt{s_{NN}} = 5.5$~TeV at the LHC this
figure is expected to increase by some  factor $\sim 2-3$.
This means that jets returned by jet definitions with a radius
parameter of, \eg, $R=0.4$, will contain background contamination of
the order of $\pi R^2 \rho \simeq$ 50 and $100-150 \GeV$ respectively.

A related, and perhaps more challenging obstacle to accurate jet
reconstruction\footnote{Note that we are considering the
  reconstruction of the jets exclusively at the {\sl particle
    level}. Detector effects can of course be relevant, and need to be
  considered in detail, but are beyond the scope of our analysis,
  which is only concerned with the removal of the background from the
  raw jet momenta. Note also that we shall be using the terms
  `(background-)subtracted momentum' and `reconstructed momentum'
  equivalently.} is due to the fluctuations both of the background level
(from event to event, but also from point to point in a single event)
and of the jet area itself: knowing the contamination level `on
average' is not sufficient to accurately reconstruct each individual
jet.

A wide range of strategies have already been advocated to reconstruct
jets in HI collisions. Just to mention them briefly, approaches to HI
jet reconstruction may involve one or more of the following measures:
choosing a jet algorithm returning jets with an area as constant as
possible; eliminating from the clustering all particles below a given
transverse momentum threshold, say $p_{t,\min} \sim 2$ GeV; measuring
the background level in a region of the detector thought to be not
affected by the hard event; or parametrising the average background
level (and fluctuations) in terms of some other measured properties of
the event, such as its centrality.

Our framework has the following characteristics:
\begin{itemize}
\item We shall restrict ourselves to IRC safe jet algorithms, to
  ensure that measurements can be meaningfully compared to
  higher-order perturbative QCD calculations.

\item The jet algorithms we use will not be limited to those yielding
  jets of regular shape.

\item Our analysis will avoid excluding small transverse momentum
  particles from the clustering. Doing so is collinear-unsafe and
  inevitably biases the reconstructed jet momenta, which must then be
  corrected using Monte Carlo simulations, which can have substantial
  uncertainties in their modelling of jet quenching and energy loss.
  Instead, we shall try to achieve a bias-free
  reconstructed jet, working with all the particles in the
  event.\footnote{
    Our framework can, of course, also accommodate the elimination of
    particles with low transverse momenta, and this might help reduce
    the dispersion of the reconstructed momentum, albeit at the
    expense of biasing it.
    Whether one prefers to reduce the dispersion or instead the bias
    depends on the specific physics analysis that one is undertaking.
    Note also that detectors may effectively introduce low-$p_t$
    cutoffs of their own. 
    These detector artefacts should not have too large an impact on
    collinear safety as long as they appear at momenta of the order of
    the hadronisation scale of QCD, \ie a few hundred MeV.
  }
\item The background level will be determined through adaptations of
  the jet-area/median procedure suggested in \cite{subtraction} and
  analysed in detail in~\cite{css}. This procedure is designed to give
  an estimate of the background that is minimally affected by the
  presence of hard jets.
\end{itemize}

%
\section {Simulation and analysis framework} \label{sec:simulation}

%
\subsection{Hard and full events} \label{sec:terminology}

In order to test the effectiveness of subtracting the underlying event in
a heavy-ion collision, and determine the quality of the jet reconstruction,
we need access to
the idealised hard jets, without the background, as a reference. 
This can be done by considering a simulated hard $pp$
event first in isolation, and then embedded in a heavy-ion event.
For clarity in the discussions, we shall adopt the following
terminology:
\begin{itemize}
\item the {\it hard event} refers to the hard $pp$ event alone,
  without a heavy-ion background;
\item the {\it full event} refers to combination of the hard event and
  the HI background (which possibly includes many semi-hard events of
  its own);
\item {\it hard jets} and {\it full jets} refers to jets from the hard
  and full events, respectively.
\end{itemize}
Note that though making the distinction between a hard and a full event is
not possible with real data, it is feasible in certain Monte Carlo
studies.
One should nevertheless be aware that the extent to which such a
distinction is physically meaningful is a question that remains open
in view of the numerous issues related to the interaction between an
energetic parton and the medium through which it travels.

If
the Monte Carlo program used to simulate heavy-ion events explicitly provides
a separation between hard events and a soft part --- \eg as does HYDJET
\cite{pyquen,hydjet} --- one can extract one of the former as the single hard event and take
the complete event as the full one. 
Alternatively, one can generate a hard event independently and embed
it in a heavy-ion event (obtained from a Monte Carlo or from real
collisions) to obtain the full event.
This second approach, which we have adopted for the bulk of results
presented here because of its greater computational efficiency, is
sensible as long as the embedded hard event is much harder than any of
the semi-hard events that tend to be present in the background.
For the transverse momenta that we consider, this condition is
typically fulfilled.
Note however that for studies in which the presence of a hard
collision is not guaranteed, \eg the evaluation of fake-jet rates,
one should use the first approach.

%
%
%
%
%
%
%

%
\subsection{Matching and quality measures} \label{sec:quality}

To measure the performance of our heavy-ion background subtraction
procedure we apply it to both hard and full events. We then take the
two hardest resulting jets in each hard event and compare them to the
corresponding subtracted jets in the full event.

To do this in practice, we need a method to match the
jets obtained from the full event to those from the hard event alone,
to make sure that we are comparing the same object. In other
analyses, this matching was typically performed in terms of the
position of the jet in the rapidity-azimuth plane, requiring that the two
jet axes not differ by more than a given $\Delta R \equiv
\sqrt{\Delta y^2 + \Delta\phi^2}$. The exact value of $\Delta R$ is an
arbitrary choice, and values between 0.1 and 0.3 have been often used.
In order to avoid the arbitrariness of the $\Delta R$ choice, we
propose a different prescription: a jet reconstructed in the full
event is considered matched to a hard jet {\it if the constituents
  common to both the hard and the full jet make up at least 50\% of
  the transverse momentum of the constituents of the hard
  jet}.\footnote{%
  Actually, in our implementation, the condition was that the common
  part should be greater than $50\%$ of the $p_t$ of the hard jet
  after UE subtraction as in section~\ref{sec:subtraction}. In
  practice, this detail has negligible impact on matching efficiencies
  and other results.
}
This definition has the advantage that for a given hard jet,
at most one full jet can satisfy this criterion, therefore avoiding having
to deal with multiple positive matchings. Another advantage is that it
automatically rejects fake matchings given by a soft jet that happens
to be close to a hard one. 

For a matched pair of jets, we shall use the notation \pthard\ and
\ptfull\ for the transverse momentum of the jet in the hard and full
event respectively, and \pthardsub\ and \ptfullsub\ for their
subtracted equivalents. The quantity that we shall mostly concentrate
on is the $p_t$ offset,
\begin{equation}\label{eq:deltapt}
  \Delta p_t \equiv \ptfullsub - \pthardsub\,,
\end{equation}
between the momentum of the background-subtracted jet in the full
event and its equivalent in the hard event. Large deviations
from zero will indicate a poor subtraction.
Events without any matched jets are simply discarded and instead
contribute to the evaluation of matching inefficiencies (see
section~\ref{sec:efficiency}).

The average of the $p_t$ shift over many events, $\langle\Delta
p_t\rangle$, is only one measure that one may examine to establish the
quality of jet reconstruction; its dispersion,
\begin{equation}
\label{eq:deltaptdispersion} 
\sigma_{\Delta p_t} \equiv \sqrt{\avg{\Delta p_t^2}-\avg{\Delta p_t}^2}\,,
\end{equation}
can be another important one. This is especially true in the case of
steeply-falling jet spectra, where this dispersion can have a large
impact on the measured cross section, necessitating delicate
deconvolutions, also known as ``unfolding''.

In this paper we shall concentrate on these two quality measures
$\avg{\Delta p_t}$ and $\sigma_{\Delta p_t}$, keeping
only the pair of jets that have been matched to one of the two hardest
(subtracted) jets in the hard event.  
Small (absolute) values of both $\avg{\Delta p_t}$ and
$\sigma_{\Delta p_t}$ will be the sign of a good subtraction. 
In practice a trade-off may exist in offset versus dispersion: which
one to optimise may depend on the specific observable one wants to
measure.%
\footnote{ Note also that $\langle\Delta p_t\rangle$ and
  $\sigma_{\Delta p_t}$ may not, in general, fully characterise the
  quality of the reconstruction, as the distribution of $\Delta p_t$
  may be non-Gaussian.}

\subsection{Monte Carlo simulations} \label{sec:MCs}

Quantifying the quality of background subtraction using Monte Carlo
simulations has  several advantages. Besides providing a practical
way of generating the hard ``signal'' separately from the soft background, one can
easily check the robustness of one's conclusions by changing the hard
jet or the background sample.

One difficulty that arises in gauging the quality of jet
reconstruction in heavy-ion collisions comes from the expectation that
parton fragmentation in a hot medium will differ from that in a
vacuum. 
This difference is often referred to as \emph{jet
  quenching}~\cite{bdmps} (for reviews, see~\cite{quenching}).
The details of jet quenching are far less well established than those
of vacuum fragmentation and can have an effect on the quality of jet
reconstruction. 
Here we shall examine the reconstruction of both unquenched and
quenched jets. 
For the latter it will be particularly important to be able to test more than
one quenching model, in order to help build confidence in our
conclusions about any reconstruction bias that may additionally exist
in the presence of quenching.

In practice, for this paper we have used both the Fortran (v1.6)
\cite{pyquen} and the C++ (v2.1) \cite{hydjet} versions of HYDJET to
generate the background. Hard jets have been generated with PYTHIA~6.4
\cite{pythia}, either running it standalone, or using the version
embedded in HYDJET v1.6. The quenching effects have been studied using
both QPYTHIA \cite{qpythia} and PYQUEN \cite{pyquen}.

%
\subsection{Jet definitions} \label{sec:jetdef}

As mentioned in the Introduction, the last few years have seen many
developments in the field of jet clustering, with the
appearance of fast implementations~\cite{fastjet_fast} of the
$k_t$~\cite{kt} and the Cambridge/Aachen~\cite{cam} sequential
recombination algorithms, and the introduction of the
SISCone~\cite{siscone} and anti-$k_t$~\cite{antikt} algorithms.

Besides these four main infrared-and-collinear-safe jet algorithms, a
number of recent papers (see \eg \cite{boosted_higgs, optimisation}),
have introduced and studied jet-cleaning techniques in order to help
the reconstruction of jets by reducing the UE
contamination. ``Filtering'' \cite{boosted_higgs} works by
reclustering each jet with a radius $R_{\rm filt}$ smaller than the
original radius $R$ and keeping only the $n_{\rm filt}$ hardest
subjets (background subtraction is applied to each of the subjets
before deciding which ones to keep).
Other similar techniques, ``trimming'' and ``pruning'', that
exploit the substructure of jets have also been introduced recently
\cite{trimming,Ellis:2009me} and, in $pp$ environments, have been
found to give benefits comparable to those of filtering.
Note that the filtering that we use here is unrelated to the Gaussian
filter approach of \cite{Lai:2008zp} (not used in this
paper as no public code is currently available).
In analogy with the systematic analysis of the performance of various
jet definitions in the kinematic reconstruction of dijet systems at
the LHC \cite{optimisation}, here we will study how different
algorithms behave in the case of HI collisions. We will use the $k_t$,
Cambridge/Aachen (C/A), and anti-$k_t$ algorithms, as well as
filtering applied to Cambridge/Aachen (C/A(filt)) with $R_{\rm
  filt}=R/2$ and $n_{\rm filt}=2$. We have not used SISCone, as its
relatively slower speed compared to the sequential recombination
algorithms makes it less suitable for a HI environment.

All the algorithms have been used through their fast implementation
available in the \fastjet package \cite{fastjet_fast,fastjet},
version~2.4.2. Additionally some features of a forthcoming FastJet
release have been used to help simplify our analyses.

%
\subsection{Background determination and subtraction} \label{sec:subtraction}

\begin{figure}[t]
  \centering
  \includegraphics[width=0.8\textwidth]{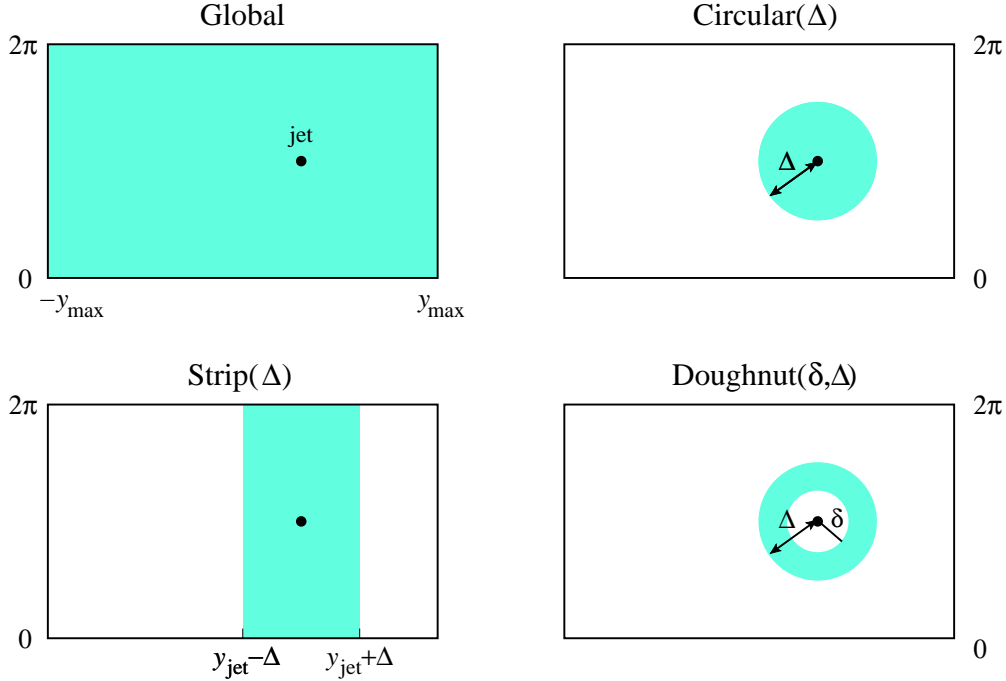}
  \caption{Graphical representation of the 4 different
    background-estimation ranges we shall
    consider: the {\em Global} range, the {\em Strip} range
    ${\cal S}_\Delta(j)$, the {\em Circular} range ${\cal C}_\Delta(j)$
    and the {\em Doughnut} range ${\cal D}_{\delta,\Delta}(j)$. The
    last three are local ranges with a position depending on the jet
    being subtracted. See the text for detailed
    definitions.}\label{fig:ranges}
\end{figure}

In order to subtract the HI background from the hard jets, we will
mainly follow the method introduced in \cite{subtraction}. 
When clustering the event we determine the 4-vector active area $A_j^\mu$
\cite{areas} of each jet $j$, as well as an estimate of the background
density $\rho$.
Then for each jet $j$, we subtract from its four-momentum the expected
background contamination:
\begin{equation}\label{eq:subtraction}
  p_j^{\mu, {\rm sub}} = p_j^\mu - \rho A_j^\mu \, .  
\end{equation}
The background transverse-momentum density per unit area, $\rho$, is
determined, event-by-event, as proposed in \cite{subtraction}. In that
paper two ways of determining $\rho$ were explored. One was to take a
global background-estimation range ${\cal R}$, covering the full
rapidity-azimuth plane up to some rapidity $y_{\max}$;
one then considered the set of ratios of transverse components of the
momentum and area 4-vectors, $p_{j,t}/A_{j,t}$, for the jets within
this range. The median of this set was used as an estimate for $\rho$:
\begin{equation}\label{eq:median}
  \rho_{{\cal R}} = \mathrm{median}\left\{\frac{p_{j,t}}{A_{j,t}}\right\}_{j \in {\cal R}},
\end{equation}
where the subscript ${\cal R}$ explicitly denotes that the background
density has been calculated using only jets in the range ${\cal R}$.
This method assumes that the background level is sufficiently constant
within ${\cal R}$. This condition is known to be violated when going
to sufficiently large rapidity, and the effect is particularly marked
in heavy ion collisions.  For this reason, it was alternatively
proposed in \cite{subtraction} to fit the mean $p_{j,t}/A_{j,t}$ as a
function of rapidity $y$ with a quadratic functional form $\rho =
\rho_0 + \rho_2 y^2$.

While working on this paper, and partly inspired by discussions with
members of the STAR collaboration (c.f.\ also ref.~\cite{star}), we came to the
conclusion that neither of these two 
methods allows one to extract sufficiently accurate values of $\rho$
in the context of HI collisions. We propose therefore a variant of the
first method, namely making the background-estimation range {\it local} and dependent on a
given jet's position. Graphical representations of a global range, as
well as of three possible local ranges that we shall use hereafter,
are given in fig.~{\ref{fig:ranges}}. More specifically, for a given
jet $j$, our three local ranges are defined as follows,\footnote{The {\em
    CircularRange} is distributed with the current \fastjet release (v2.4.2) 
  and the {\em StripRange} can be simulated from the default {\em
    RangeDefinition}. A more systematic approach to local ranges,
  including the {\em DoughnutRange}, will be available in the
  forthcoming \fastjet release.}
\begin{itemize}
\item the {\it Strip} range, ${\cal S}_{\Delta}(j)$, includes the jets
  $j'$ satisfying $|y_{j'}-y_j|<\Delta$,
\item the {\it Circular} range, ${\cal C}_{\Delta}(j)$, includes the
  jets $j'$ satisfying
  $\sqrt{(y_{j'}\!-\!y_j)^2+(\phi_{j'}\!-\!\phi_j)^2} < \Delta$,
\item the {\it Doughnut} range, ${\cal D}_{\delta,\Delta}(j)$, includes
  the jets $j'$ satisfying $\delta <
  \sqrt{(y_{j'}\!-\!y_j)^2+(\phi_{j'}\!-\!\phi_j)^2} < \Delta$.
\end{itemize}
Using ${\cal R}(j)$ to denote any local range around jet $j$, the
background density will then depend on the jet being subtracted and
will be estimated using
\begin{equation}\label{eq:local_median}
  \rho_{{\cal R}(j)} = \mathrm{median}\left\{\frac{p_{j'\!,t}}{A_{j'\!,t}}\right\}_{j' \in {\cal R}(j)}.
\end{equation}
It is important to be aware that the estimate of the background
density $\rho$ is just an input to eq.~(\ref{eq:subtraction}) and that the
jet definition that is used for the computation of $\rho$ in
eqs.~(\ref{eq:median}) or (\ref{eq:local_median}) 
can be different from  the jet definition used to
obtain the ``physical'' jets and subtract them in
eq.~(\ref{eq:subtraction}). 
The only two recommended jet algorithms 
for the task of determining $\rho$ are the $k_t$ and C/A
algorithms (see \cite{subtraction}); others are not ideal because they
return many jets with very small areas, which distorts the median
procedure. 
The radius, $R_\rho$, used in the jet definition for determining
$\rho$ can also differ from that, $R$, used to find the jets.

When using a local range, the underlying idea is to limit the sensitivity 
to
the long-range variations of the background density, by using only the
jets in the vicinity of the jet we want to subtract.
In practice, a compromise needs to be found between choosing a range 
small enough to get a valid local estimation,  but also large enough
to contain a sufficiently large number of background (soft) jets 
for the estimation of the median to be reliable.
Two effects need to be considered: (1) statistical fluctuations in the
estimation of the background and (2) biases due to the presence of a hard
jet in the region used to estimate the background.
%
%
%
%
\begin{enumerate}
\item If we require that the dispersion in the reconstructed jet $p_t$
  coming from the statistical fluctuations in the estimation of the
  background does not amount to more than a fraction $\epsilon$ of the
  overall $\sigma_{\Delta p_t}$, then as discussed in
  appendix~\ref{app:minrange-fluct} the range needs to cover an area $A_{\cal
    R}$ such that
  \begin{equation}
    \label{eq:Ar-lower-lim}
    A_{\cal R} \gtrsim \frac{\pi^2 R^2}{4\epsilon}\,.
  \end{equation}
  Taking $\epsilon = 0.1$ and $R=0.4$ this corresponds to an area
  $A_{\cal R} \gtrsim 25 R^2 \simeq 4$.
  For the applications below, when clustering with radius $R$, we have
  used the rapidity-strip ranges ${\cal S}_{2R}$ and ${\cal S}_{3R}$,
  the circular range ${\cal C}_{3R}$ and the doughnut range ${\cal
    D}_{R,3R}$. One can check that their areas are compatible with the $25
  R^2$ lower-limit estimated above.

\item The bias in the estimate of $\rho$ due to the presence of $n_b$
  hard jets in the range is given roughly by
  \begin{equation}
    \label{eq:hard_median_offset}
    \langle \Delta \rho \rangle \simeq 1.8 \sigma R_\rho
    \frac{n_b}{A_{\cal R}}\,,
  \end{equation}
  as discussed in \cite{css} and appendix~\ref{app:minrange-bias}.
  The ensuing bias on the $p_t$ can be estimated as $\langle \Delta
  \rho \rangle \pi R^2$ (for anti-$k_t$ jets). 
  For $R=0.4$, $R_\rho = 0.5$, $A_{\cal R} \simeq 4$ and $n_b=1$, the
  bias in the reconstructed jet $p_t$ is $\simeq 0.1 \sigma$. 
  Given $\sigma$ in the $10-20\GeV$ range (as we will find in
  section~\ref{sec:results}) this corresponds to a $1-2\GeV$ bias.
  In order to eliminate this small bias, we will often choose to
  exclude the two hardest jets in each event when determining of
  $\rho$.\footnote{We deliberately choose to exclude the two hardest
    jets in the event, not simply the two hardest in the range. Note,
    however, that for realistic situations with limited acceptance,
    only one jet may be within the acceptance, in which case the
    exclusion of a single jet might be more appropriate. Excluding a
    second one does not affect the result significantly, so it is
    perhaps a good idea to use the same procedure regardless of any
    acceptance-related considerations.}
  
\end{enumerate}
A third potential bias discussed in ref.~\cite{css} is that of
underestimating the background when using too small a value for
$R_\rho$. 
Given the high density of particles in HI collisions, this will 
generally not be an issue as long as $R_\rho\sim 0.5$.

%
\section{Results} \label{sec:results}

\begin{figure}
\centerline{
\includegraphics[angle=270,width=0.5\textwidth]{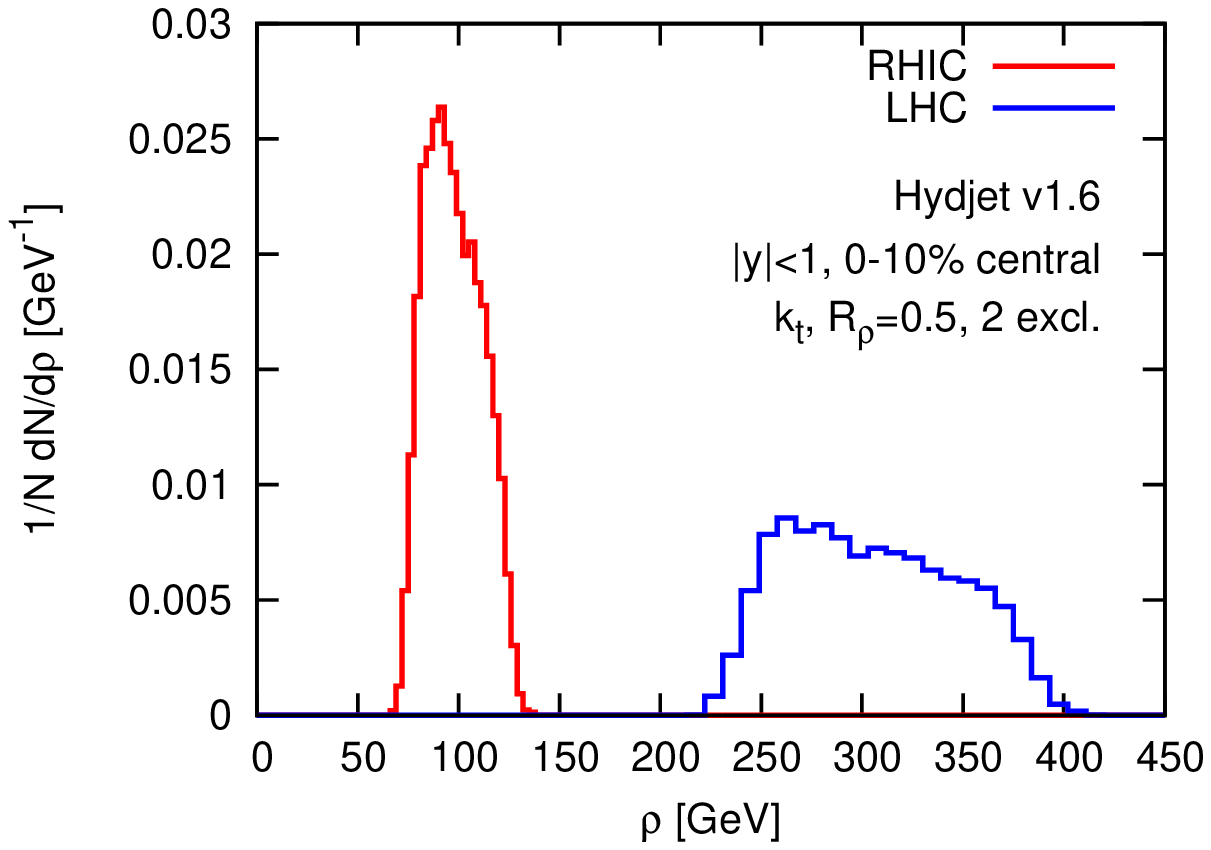}
\includegraphics[angle=270,width=0.5\textwidth]{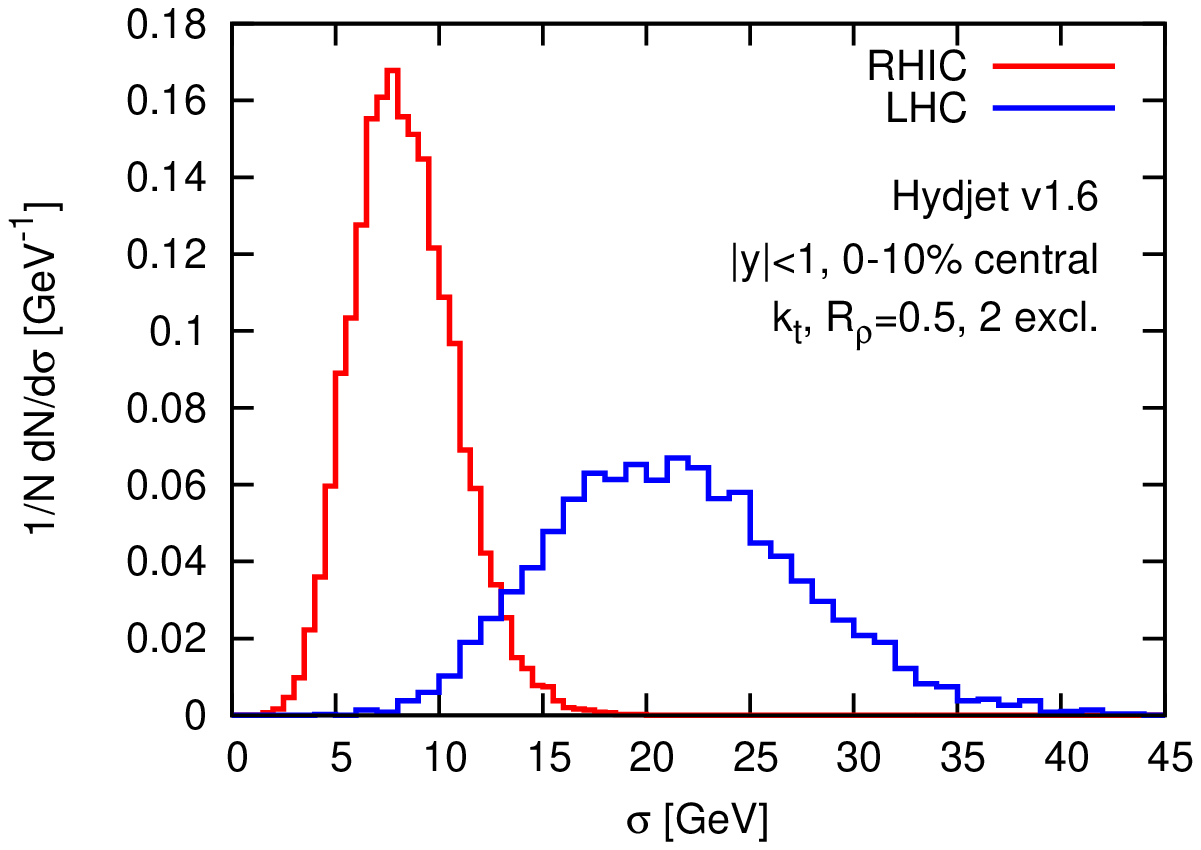}
}
\caption{\label{fig:bkg_distrib}
  Distribution of the background density $\rho$ per unit area (left)
  and its intra-event fluctuations $\sigma$ (right). It has been
  obtained from 5000 HYDJET
  events with RHIC (\AuAu, $\sqrt{s_{NN}} = 200\GeV$) and LHC
  (\PbPb, $\sqrt{s_{NN}} = 5.5\TeV$) kinematics. The background properties
  have been estimated using the techniques presented in section
  \ref{sec:subtraction}, using the $k_t$ algorithm with $R_\rho=0.5$, and
  keeping only the jets with $|y|<1$ (excluding the two hardest).}
\end{figure}

As mentioned already in section~\ref{sec:MCs}, for the simulation of
the events, we have used PYTHIA 6.4 \cite{pythia} to generate the
unquenched hard jets; the background events are generated using HYDJET
v1.6 \cite{pyquen} with 0-10\% centrality\footnote{
  The events have been generated using the following HYDJET~v1.6
  program parameters: for RHIC, $\texttt{nh}=9000$,
  $\texttt{ylfl}=3.5$, $\texttt{ytfl}=1.3$ and
  \texttt{ptmin}$=2.6\GeV$;
  for LHC, $\texttt{nh}=30000$, $\texttt{ylfl}=4$, $\texttt{ytfl}=1.5$
  and $\texttt{ptmin}=10\GeV$. In both cases quenching effects are
  turned on in HYDJET, $\texttt{nhsel}=2$, even when they are not
  included for the embedded $pp$ event. The corresponding PYQUEN
  parameters we have 
  used are $\texttt{ienglu}=0$,  $\texttt{ianglu}=0$,
  $\texttt{T0}=1.0\GeV$, $\texttt{tau0}=0.1$~fm and $\texttt{nf}=0$.
  \label{foot:pyquen}}.
Our setup for the background leads, for RHIC (\AuAu, $\sqrt{s_{NN}} =
200\GeV$), to an average background density per unit area at central
rapidity of $\langle\rho\rangle \simeq 99$~GeV with average
fluctuations in a single event of $\langle\sigma\rangle \simeq 8$~GeV
and event-to-event fluctuations $\sigma_\rho \equiv \sqrt{\langle
  \rho^2\rangle - \langle \rho\rangle^2} \simeq 14$~GeV.  
For the LHC (\PbPb, $\sqrt{s_{NN}} = 5.5\TeV$) the corresponding
values are $\langle\rho\rangle \simeq 310$~GeV, $\langle\sigma\rangle
\simeq 20$~GeV and event-to-event fluctuations $\sigma_\rho \simeq
45$~GeV.
Fig.~\ref{fig:bkg_distrib} shows the distributions obtained from the
simulations for $\rho$ and $\sigma$.

In the case of the RHIC simulation, the result for $\avg{\rho}$ is
somewhat higher than the experimental value of $75\GeV$ quoted by the STAR
collaboration \cite{star}, however this is probably in part due to
limited tracking efficiencies at low $p_t$ at STAR,\footnote{We thank
  Helen Caines for discussions on this point.} and explicit
STAR~\cite{Adams:2004cb} and PHENIX~\cite{Adler:2004zn} results for
$dE_t/d\eta$ correspond to somewhat higher $\rho$ values, about
$90\GeV$.
The multiplicity of charged particles ($dN_{ch}/d\eta \simeq 660$ for
$\eta=0$ and 0--6\% centrality)
and the pion $p_t$ spectrum in our simulation are
sensible compared to experimental measurements at RHIC
\cite{dnchdy,dnchdpt,dn0dpt}.
For the LHC our charged particle multiplicity is $dN_{ch}/d\eta \simeq
1600$ for $\eta=0$ and 0--10\% centrality, which is comparable to many
of the predictions reviewed in fig.~7 of \cite{Armesto:2009ug}.

An independent control analysis has also been performed with
HYDJET++~2.1~\cite{hydjet} (with default parameters) for the
background. The results at RHIC are similar, while for LHC the
comparison is difficult because the default tune of HYDJET++~2.1
predicts a much higher multiplicity, $dN_{ch}/d\eta \simeq 2800$ for
$\eta=0$ and 0--10\% centrality.

Most of the results of this section will be obtained without
quenching, though in section~\ref{sec:quenching} we will also consider
the impact on our conclusions of the PYQUEN~\cite{pyquen} and
QPYTHIA~\cite{qpythia} simulations of quenching effects.

%
%
%
%
%
%
%
%
%
%

For the results presented below, we have employed a selection cut of
$|y|\le y_{\rm max}$ on the jets with
$y_{\rm max}=1$ for RHIC and $y_{\rm max}=2.4$ for the
LHC.\footnote{This corresponds roughly to the central region for the
  ATLAS and CMS detectors. For ALICE, the acceptance is more
  limited~\cite{AliceHija,Alice0910}. Some adaptation of our method
  will be needed for estimating $\rho$ in that case, in order for
  information to be derived from jets near the edge of the acceptance
  and thus bring the available area close to the ideal requirements
  set out in appendix~\ref{sec:minimal-range}.
  Note that we also use particles beyond $y_{\max}$ in the jet
  clustering and apply the acceptance cut only to the resulting jets.
}
We only consider full jets that are matched to one of the two hardest
jets in the hard event.
The computation of the jet areas in \fastjet, needed both for the
subtraction and the background estimation, has been performed using
active areas, with ghosts up to $y_{\rm max}+1.8$, a single repetition
and a ghost area of 0.01.%
\footnote{The active area~\cite{areas} is the natural choice for
  subtraction as it mimics the uniform soft background. 
  We also use the ``explicit ghosts'' option of \fastjet, which gives
  a better computation of the empty area in sparse events. For the C/A
  algorithm with filtering, explicit ghosts also allow for subtraction
  of each individual subjet before selecting the two hardest subjets.
  Finally, note that since we are mostly dealing with
  high-multiplicity events, the difference between active and passive
  areas is negligible, and we could in some cases also have used the
  latter (\eg to limit certain speed and memory issues if we had used
  SISCone).
}
%
%
%
%
%
%
%
%
%
%
%
%
%
%
%
The determination of the background density $\rho$ has been performed
using the $k_t$ algorithm with $R_\rho=0.5$. Though the estimate of
the background depends on $R_{\rho}$ \cite{css}, we have observed that
choices between 0.3 and 0.5 lead to very similar results (\eg
differing by at most a few hundred MeV at RHIC).


With this setup, we have studied the various ranges presented in
section \ref{sec:subtraction}, with the jet algorithms from section
\ref{sec:jetdef}. In all cases, we have taken the radius parameter
$R=0.4$. We have adopted this value as it is the largest currently
used at RHIC. Note that the effect of the background fluctuations of
the jet energy resolution increases linearly with $R$, disfavouring
significantly larger choices. 
On the other hand, too small a choice of $R$ may lead to excessive
sensitivity to the details of parton fragmentation, hadronisation and
detector granularity.
%
%

%
%
%
%
%
%
%
%
%
%
%

%
\subsection{Matching efficiency}\label{sec:efficiency}

\begin{figure}
\centerline{
\includegraphics[angle=270,width=0.5\textwidth]{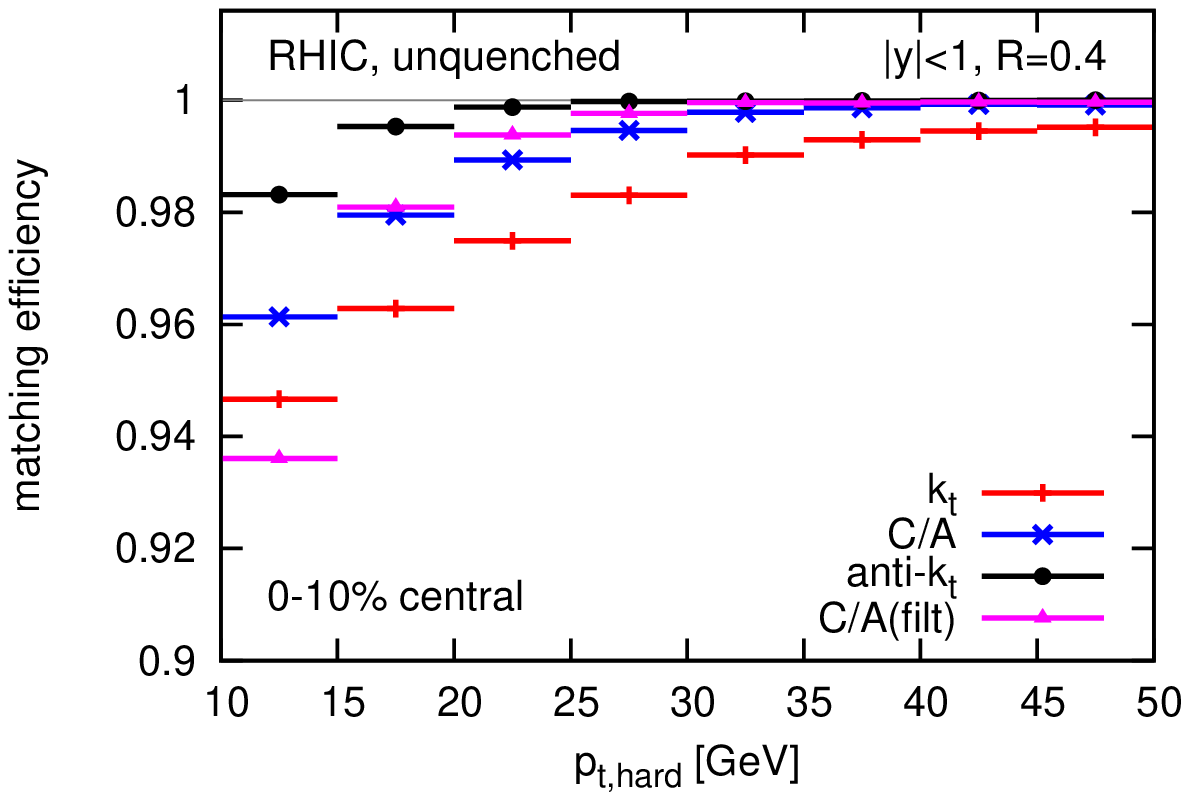}
\includegraphics[angle=270,width=0.5\textwidth]{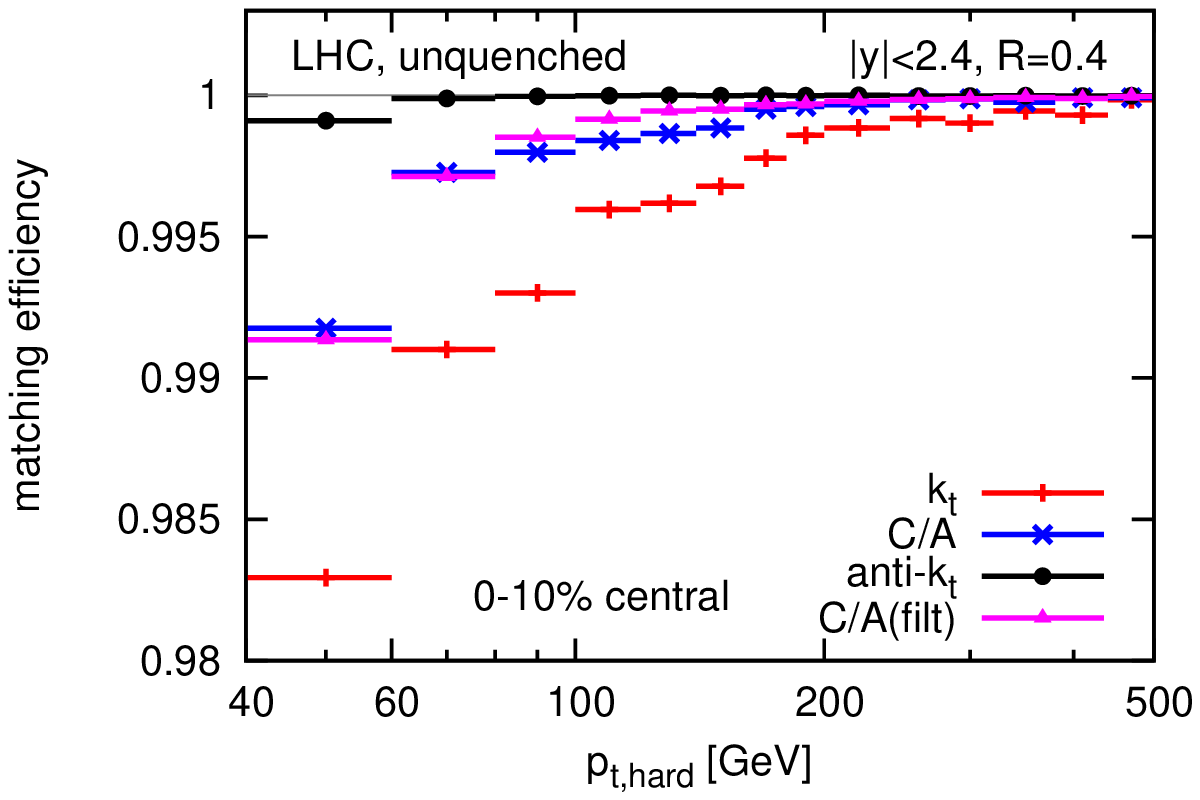}
}
\caption{\label{fig:efficiency} Matching efficiency for reconstructed
  jets as a function of the jet $p_t$. Left: RHIC, right: LHC. These
  results are independent of the choice of background-subtraction
  range in the heavy-ion events, since background subtraction does not
  enter into the matching criterion. 
  Here and in later figures, the label ``unquenched'' refers to
  the embedded $pp$ event; the background is always simulated
  including quenching.
}
\end{figure}

Let us start the presentation of our results with a brief discussion
of the efficiency of reconstructing jets in the medium. As explained
in section \ref{sec:quality}, the jets in the medium are matched to a
``bare'' hard jet when their common particle content accounts for at
least 50\% of the latter's transverse momentum.

The matching efficiencies we observe depend to some extent on the
details of the Monte Carlo used for the background so our intention is
just to illustrate the typical behaviour we observe and highlight that
these efficiencies tend to be large. We observe from
fig. \ref{fig:efficiency} that we successfully match at least 95\% of
the jets above $p_t \simeq 15$ GeV at RHIC, and at least 99\% of the
jets above $p_t \simeq 60$ GeV at the LHC. It is also interesting to
notice that the anti-$k_t$ algorithm performs best, likely as a
consequence of its `rigidity', namely the fact that anti-$k_t$ jets tend 
to have the same (circular) shape, independently of the soft-particles
that are present.

%
\subsection{Choice of background-estimation range}\label{sec:choice_range}

\begin{figure}[t]
\centerline{
\includegraphics[angle=270,width=0.5\textwidth]{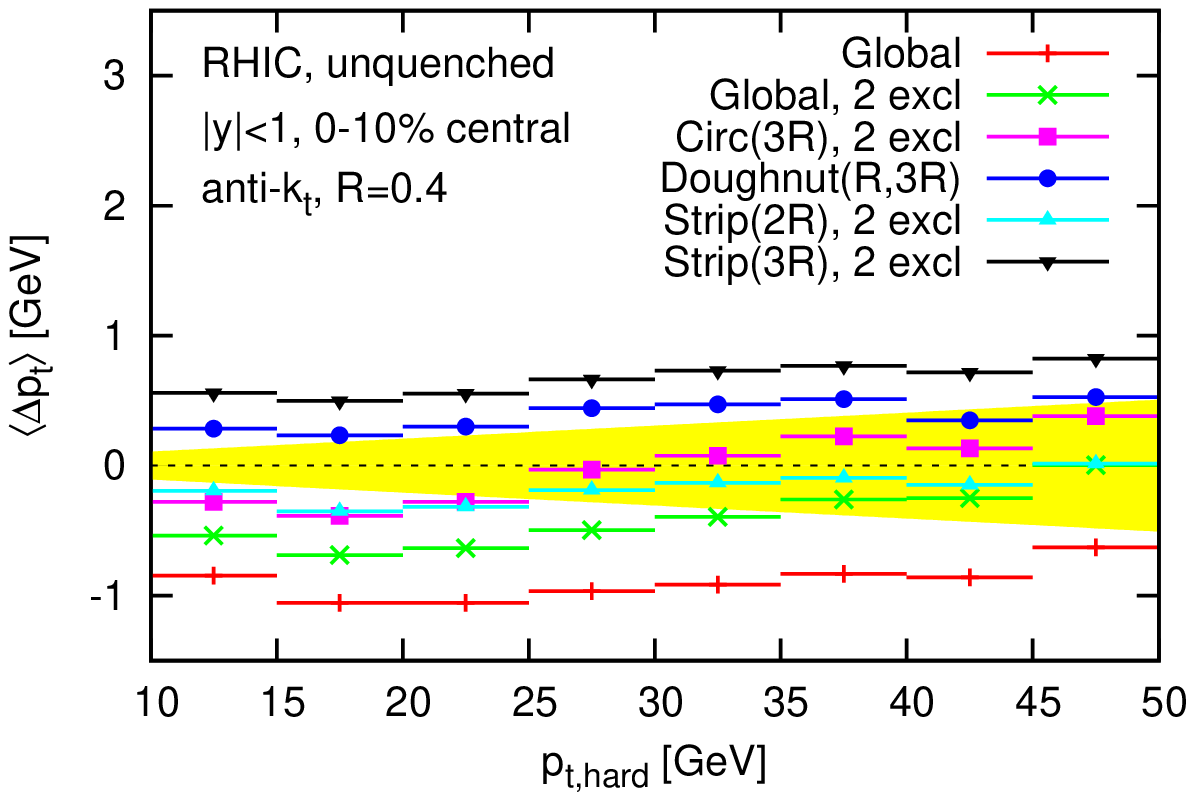}
\includegraphics[angle=270,width=0.5\textwidth]{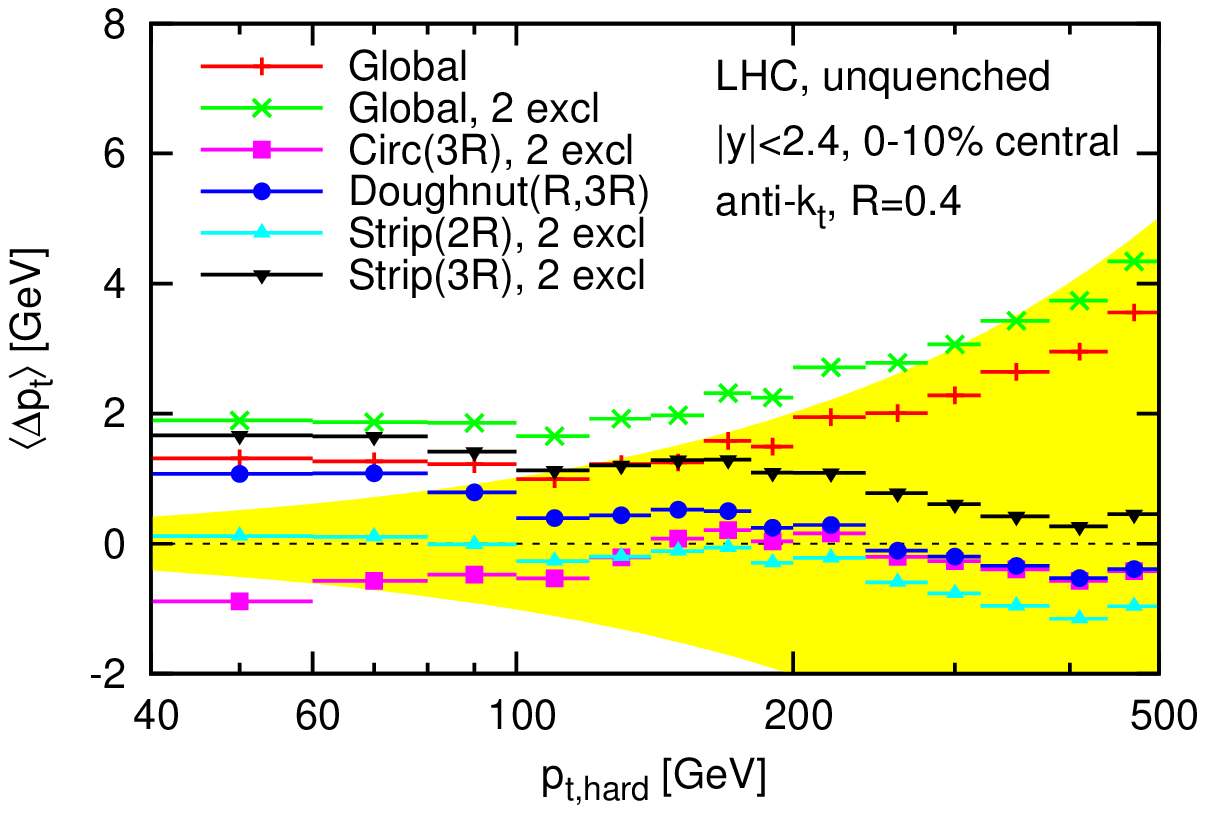}
}
\caption{\label{fig:ptshift_ranges} Effect of the choice of range on
  the average $p_t$ shift, $\Delta p_t$, as defined in
  eq.~(\ref{eq:deltapt}). Left: RHIC, right: LHC. In this figure and
  those that follow, the yellow band corresponds to 1\% of the $p_t$
  of the hard jet.}
\end{figure}

We now turn to the results concerning the measurement of the
background density and the reconstruction of the jet transverse
momentum. We first concentrate on the impact of the choice of a local
range and/or of the exclusion of the two hardest jets when determining
$\rho$.\footnote{Independently of the choice made for the full event,
  we always use a global range up to $|y| = y_{\max}$ for the
  determination of $\rho$ in the hard event, without exclusion of any
  jets. This ensures that the reference jet $p_t$ is always kept the
  same. The impact of subtraction in the hard event is in any case
  small, so the particular choice of range is not critical.}

In fig.~\ref{fig:ptshift_ranges} we show the average shift
$\avg{\Delta p_t}$ for the list of ranges mentioned in section
\ref{sec:subtraction}. The results presented here have been obtained
with the anti-$k_t$ algorithm with $R=0.4$, but the differences among
the various range choices have been seen to be similar with other jet
definitions.  The label ``2 excl'' means that the two hardest jets in
the event have been excluded from the estimation of the background. 
We have found that this improves the precision of the subtractions
whenever expected, \ie for all choices of range except the doughnut
range, where its central hole already acts similarly to the exclusion
of the hardest jets.
To keep the figure reasonably
readable, we have only explicitly shown the effect of removing the two
hardest jets for the global range.
The change of 0.4--0.6~GeV (both for RHIC and the LHC) is
in reasonable agreement with the analytic estimate of about 0.6 GeV
for RHIC and the LHC obtained from $\avg{\Delta p_t}=\pi R^2
\avg{\Delta\rho}$ with $\avg{\Delta\rho}$ calculated using
eq.~(\ref{eq:hard_median_offset}).
Note that at LHC the exclusion of the two hardest jets for the global
range appears to worsen the subtraction, however what is really
happening is that the removal of the two hardest jets exacerbates a
deficiency of the global range, namely the fact that its broad
rapidity coverage causes it to underestimate $\rho$, leading to a
positive net $\avg{\Delta\rho}$.

Other features that can be understood qualitatively include for
example the differences between the two strip and the global (2~excl)
range for RHIC: while the rapidity width of the global range lies
in between that of the two strip ranges, the global range gives a
lower $\avg{\Delta p_t}$ than both, corresponding to a larger $\rho$
estimate, which is reasonable because the global range is centred on
$y=0$, whereas the strip ranges are mostly centred at larger
rapidities where the background is lower.

The main result of the analysis of fig.~\ref{fig:ptshift_ranges} is
the observation that all choices of a local range lead to a small
residual $\Delta p_t$ offset: the background subtraction typically
leaves a $|\avg{\Delta p_t}| \lesssim 1\GeV$ at both RHIC and LHC, \ie
better than 1-2\% accuracy over much of the $p_t$ range of interest.
It is not clear, within this level of accuracy, if one range is to be
preferred to another, nor is it always easy to identify the precise
origins of the observed differences between various
ranges.\footnote{Furthermore, the differences may also be modified by
  jet-medium interactions. }
Another way of viewing this is that the observed differences between
the various choices give an estimate of the residual subtraction error
due to possible misestimation of $\rho$.
For our particular analysis, at RHIC this comment also applies to the
choice of the global range (with the exclusion of the two hardest jets
in the event). This is a consequence of the limited rapidity
acceptance, which effectively turns the global range into a local one,
a situation that does not hold for larger rapidity acceptances, as we
have seen for the LHC results.
In what follows we will use the Doughnut($R,3R$) choice, since it
provides a good compromise between simplicity and effectiveness.

%
%
%
%
%

%
\subsection{Choice of algorithm}\label{sec:choice_alg}

\begin{figure}
  \centering
  \includegraphics[height=0.9\textwidth,angle=-90]{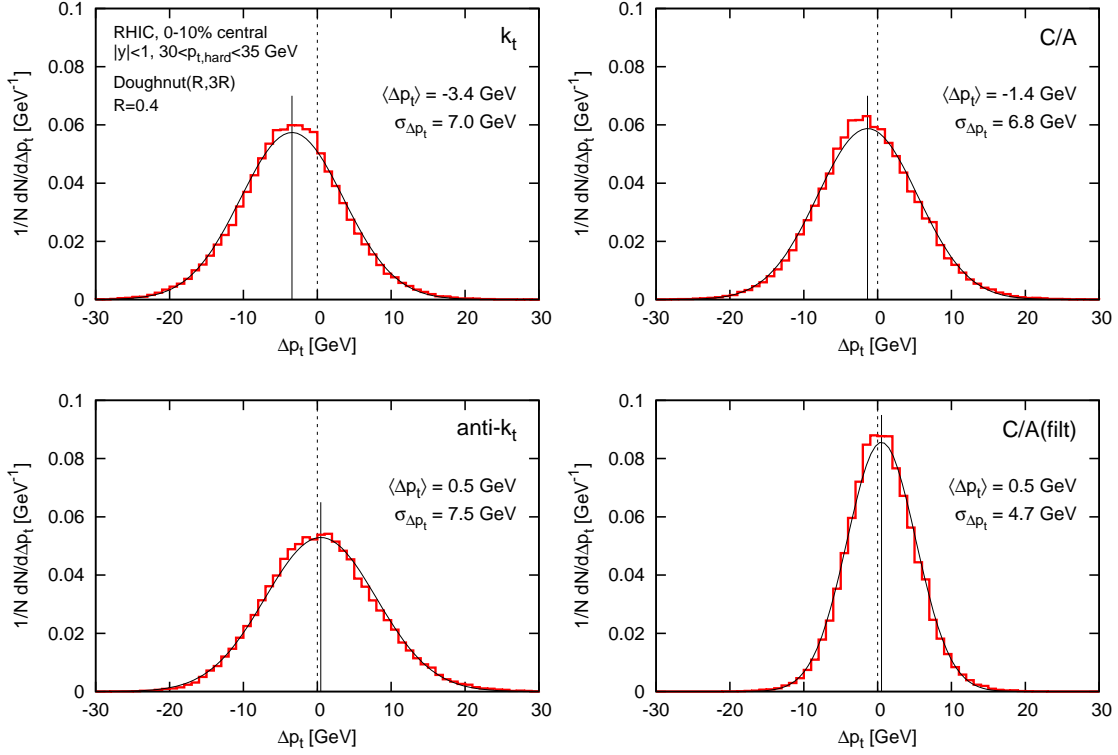}
  \caption{Distribution of $\Delta p_t$ (red histograms) for each of
    our 4 jet algorithms, together with a Gaussian (black curve) whose
    mean (solid vertical line) and dispersion are equal to
    $\avg{\Delta p_t}$ and $\sigma_{\Delta p_t}$ respectively.}
  \label{fig:deltapt-dist}
\end{figure}

The next potential systematic effect that we consider is the choice of
the jet algorithm used for the clustering.\footnote{Recall that in all
  cases, the $k_t$ algorithm with $R_\rho=0.5$ is used for the
  estimation of the background.}
Fig.~\ref{fig:deltapt-dist} shows the distribution of $\Delta p_t$ for each
of our four choices of jet algorithm, $k_t$, C/A, anti-$k_t$ and
C/A(filt), given for RHIC collisions and a specific bin of the hard
jets' transverse momenta, $30 < p_{t,\hard} < 35 \GeV$.
One sees significant differences between the different algorithms.
One also observes that Gaussians with mean and dispersion set equal to
$\avg{\Delta p_t}$ and $\sigma_{\Delta p_t}$ provide a fair
description of the full histograms. 
This validates our decision to concentrate on $\avg{\Delta p_t}$ and
$\sigma_{\Delta p_t}$ as quality measures. 
One should nevertheless be aware that in the region of high $|\Delta
p_t|$ there are deviations from perfect Gaussianity, which are more
visible if one replicates fig.~\ref{fig:deltapt-dist} with a
logarithmic vertical scale (not shown, for brevity).

\subsubsection{Average $\Delta p_t$ }
\label{sec:av-dlt-pt}

\begin{figure}
\centerline{
\includegraphics[angle=270,width=0.5\textwidth]{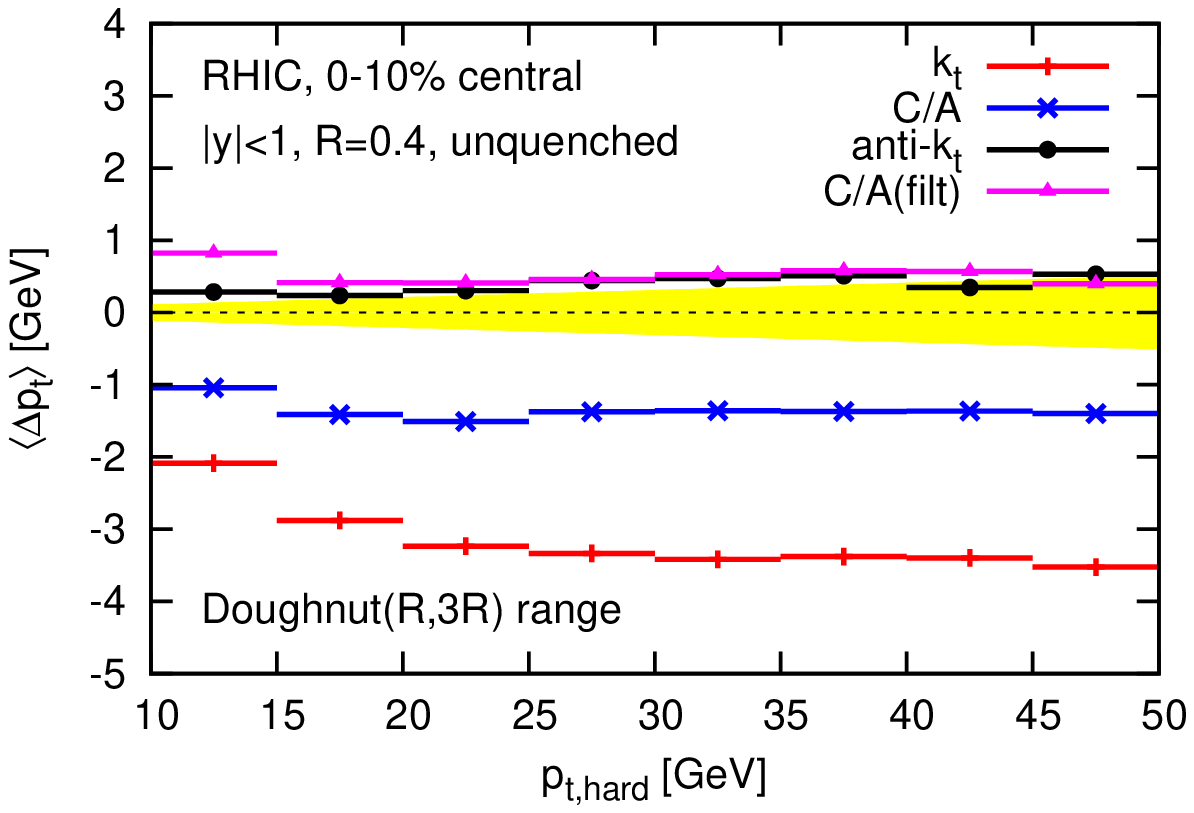}
\includegraphics[angle=270,width=0.5\textwidth]{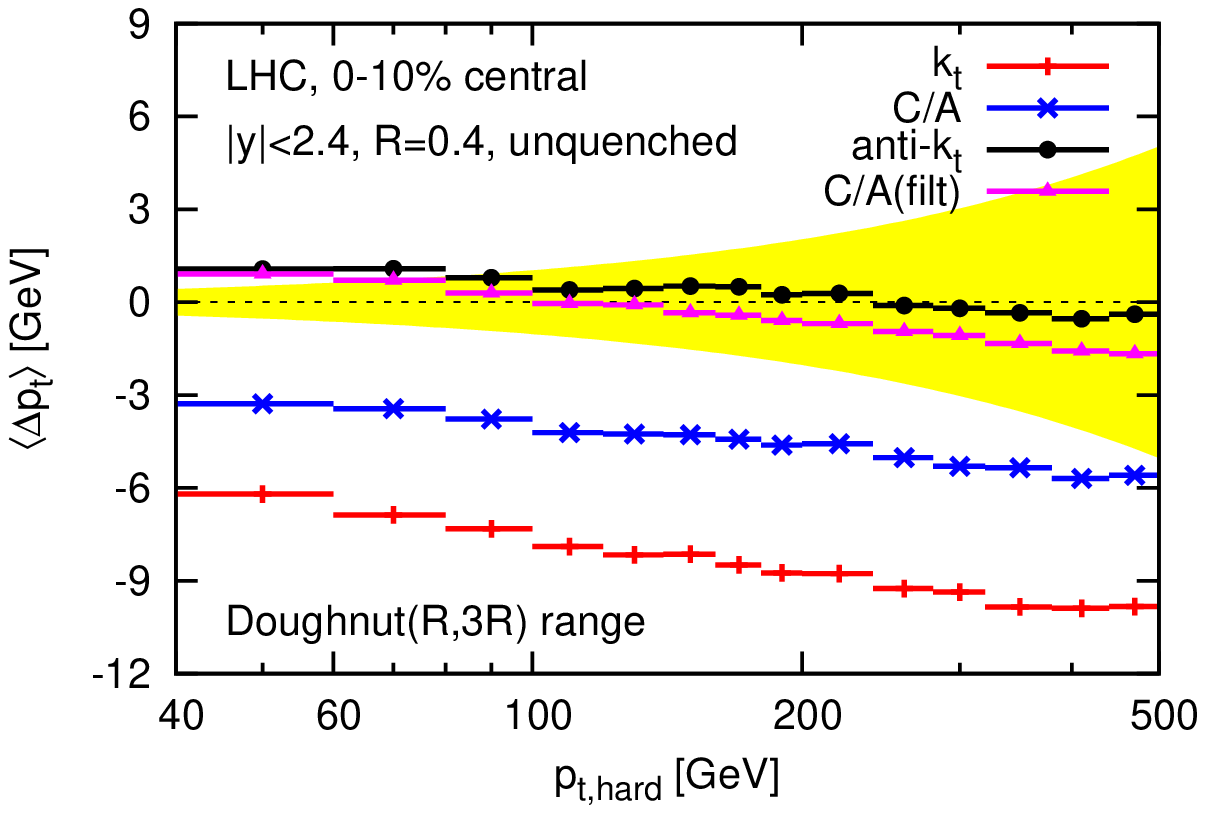}
}
\caption{\label{fig:ptshift} Average shift $\avg{\Delta p_t}$, as a
  function of $p_{t,\hard}$, shown for RHIC (left) and the LHC
  (right).}
\end{figure}

The first observable we analyse is the average $p_t$ shift. We show in
fig.~\ref{fig:ptshift} the $\langle\Delta p_t\rangle$ results for the
four algorithms listed in section \ref{sec:jetdef}, as a function of
$p_{t,\hard}$.  
We use the doughnut range to estimate the background.
The first observation is that, while the anti-$k_t$ and
C/A(filt) algorithms have a small residual $\avg{\Delta p_t}$, the C/A
and $k_t$ algorithms display significant offsets.
The reason for the large offsets of $k_t$ and C/A is well understood,
related to an effect known as {\em back-reaction}~\cite{areas}. This
is the fact that the addition of a soft background can alter the
clustering of the particles of the hard event: some of the
constituents of a jet in the hard event can be gained by or lost from
the jet when clustering the event with the additional background of
the full event. 
This happens, of course, on top of the simple background contamination
that adds background particles to the hard jet. Even if this latter
contamination is subtracted exactly, the reconstructed $p_t$ will
still differ from that of the original hard jet as a consequence of
the back-reaction.

\begin{figure}
\centerline{
\includegraphics[angle=270,width=0.5\textwidth]{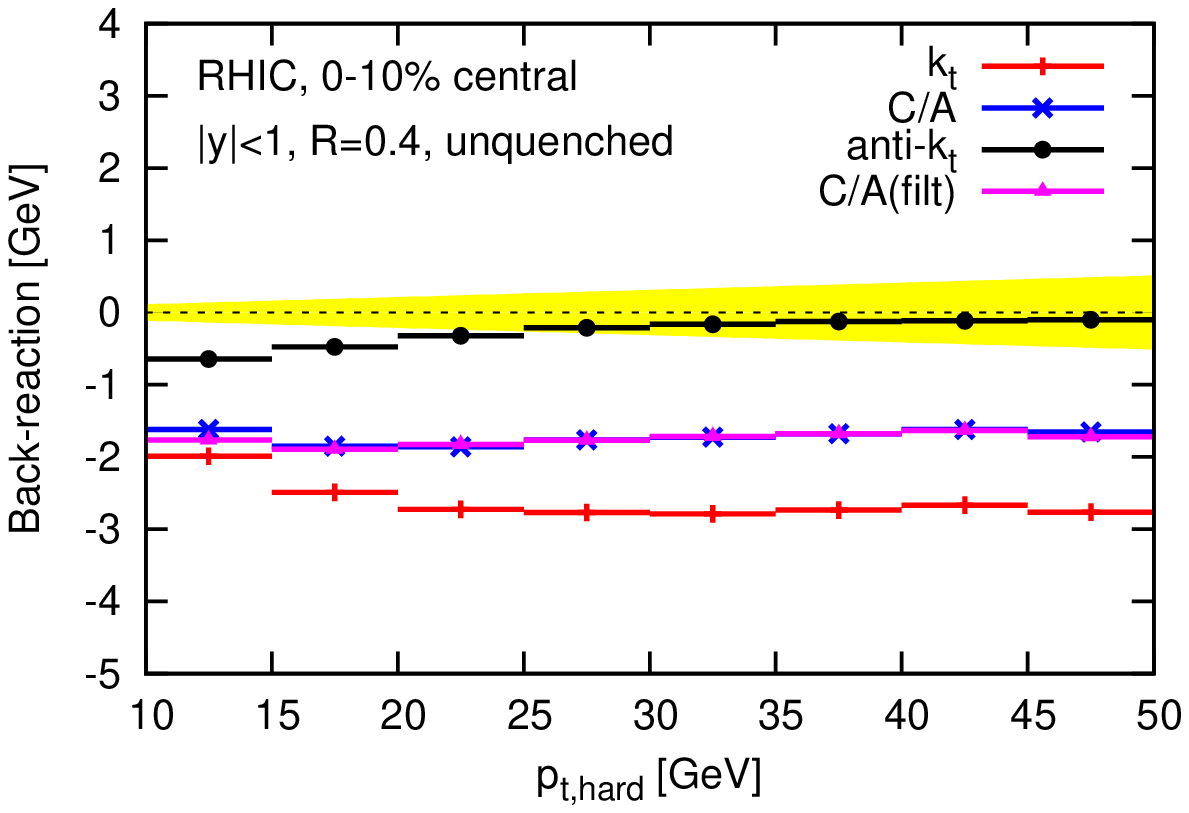}
\includegraphics[angle=270,width=0.5\textwidth]{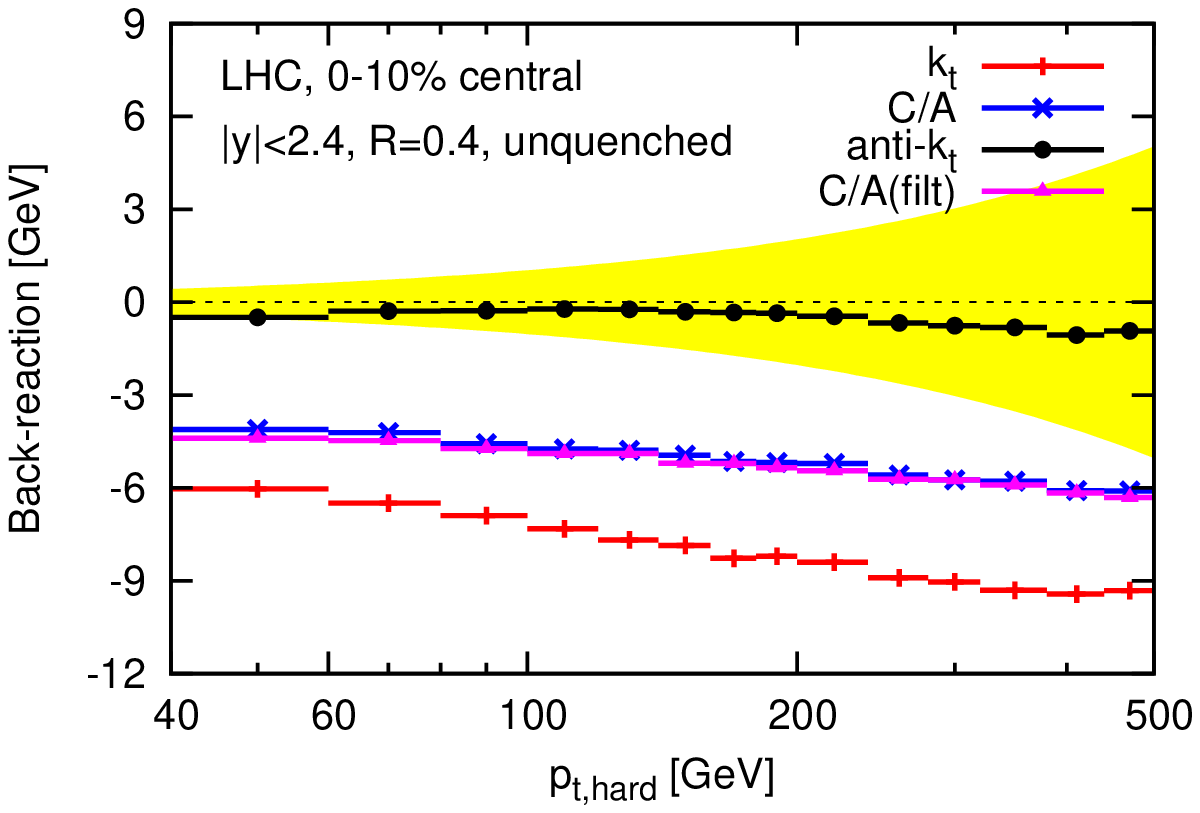}
}
\caption{\label{fig:ptshift_br} Contribution to $\avg{\Delta p_t}$ due
  to back-reaction. Note that these results are independent of the
  range used for estimating $\rho$ in the heavy-ion event. Left: RHIC,
  right: LHC.}
\end{figure}

The effect of the back-reaction can be studied in detail, since in
Monte Carlo simulations it is possible to identify which hard-event
constituents are present in a given jet before and after inclusion of
the background particles in the clustering.
The average $p_t$ shift due to back-reaction 
can be seen in fig.~\ref{fig:ptshift_br} for the different jet algorithms.
As expected \cite{areas,antikt}, it is largest for $k_t$, and smallest (almost
zero, in fact) for anti-$k_t$.
By comparing fig.~\ref{fig:ptshift_br} and fig.~\ref{fig:ptshift} one
can readily explain the difference between the $\avg{\Delta p_t}$
offsets of the various algorithms in terms of their back-reaction.
The rigidity (and hence small back-reaction) of the anti-$k_t$ jets
manifestly gives almost bias-free reconstructed jets, while the large
back-reaction effects of the $k_t$ algorithm and, to a smaller extent,
of the C/A algorithm translates into a worse performance in terms of
average shift.
The $p_t$ dependence of the back-reaction is weak. This is expected
based on the interplay between the $\ln \ln p_t$ dependence found in
\cite{areas} and the evolution with $p_t$ of the relative fractions of
quark and gluon jets.

The case of the C/A(filt) algorithm is more complex: its small net
offset, comparable to that of the anti-$k_t$ algorithm, appears to be
due to a fortuitous compensation between an under-subtraction of the
background and a negative back-reaction.
The negative back-reaction is very similar to that of C/A without
filtering, while the under-subtraction is related to the fact that the
selection of the hardest subjets introduces a bias towards
positive fluctuations of the background. This effect is discussed in
Appendix~\ref{app:filtbias}, where we obtain the following estimate
for the average $p_t$ shift (specifically for $R_{\rm filt}=R/2$):
\begin{equation}
  \label{eq:filt-bias-main-text}
\left\langle (\Delta p_t)_{\rm filt}\right\rangle
  \simeq 0.56\,R \sigma,
\end{equation}
yielding an average bias of 2~GeV for RHIC and 4.5~GeV at the LHC,
which are both in good agreement with the differences observed between
C/A with and without filtering in fig.~\ref{fig:ptshift}.
Note that while the bias in eq.~(\ref{eq:filt-bias-main-text}) is
proportional to $\sigma$, the back-reaction bias is instead mainly
proportional to $\rho$~\cite{areas}.
It is because of these different proportionalities that the
cancellation between the two effects should be considered as
fortuitous.
Since it also depends on the substructure of the jet, one may also
expect that the cancellation that we see here could break down in the
presence of quenching.

\subsubsection{Dispersion of $\Delta p_t$ }
\label{sec:dispersion}

\begin{figure}
\centerline{
\includegraphics[angle=270,width=0.5\textwidth]{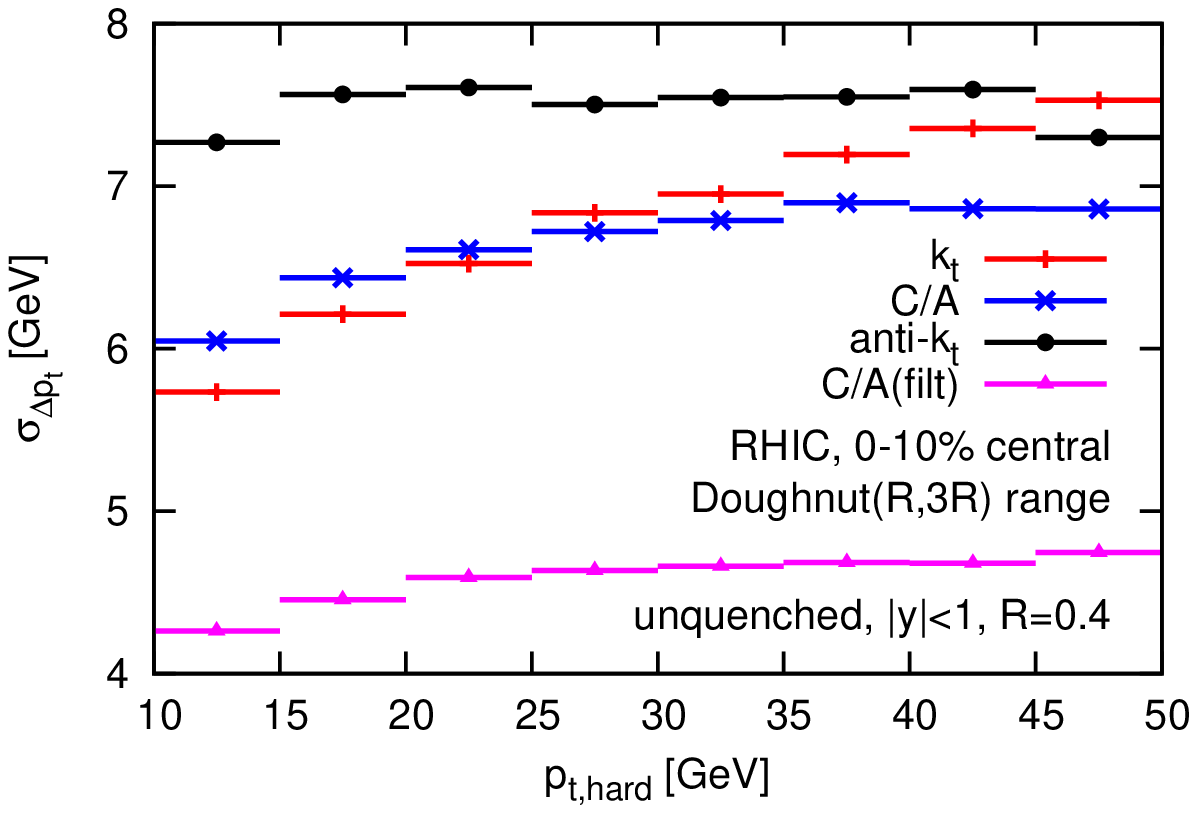}
\includegraphics[angle=270,width=0.5\textwidth]{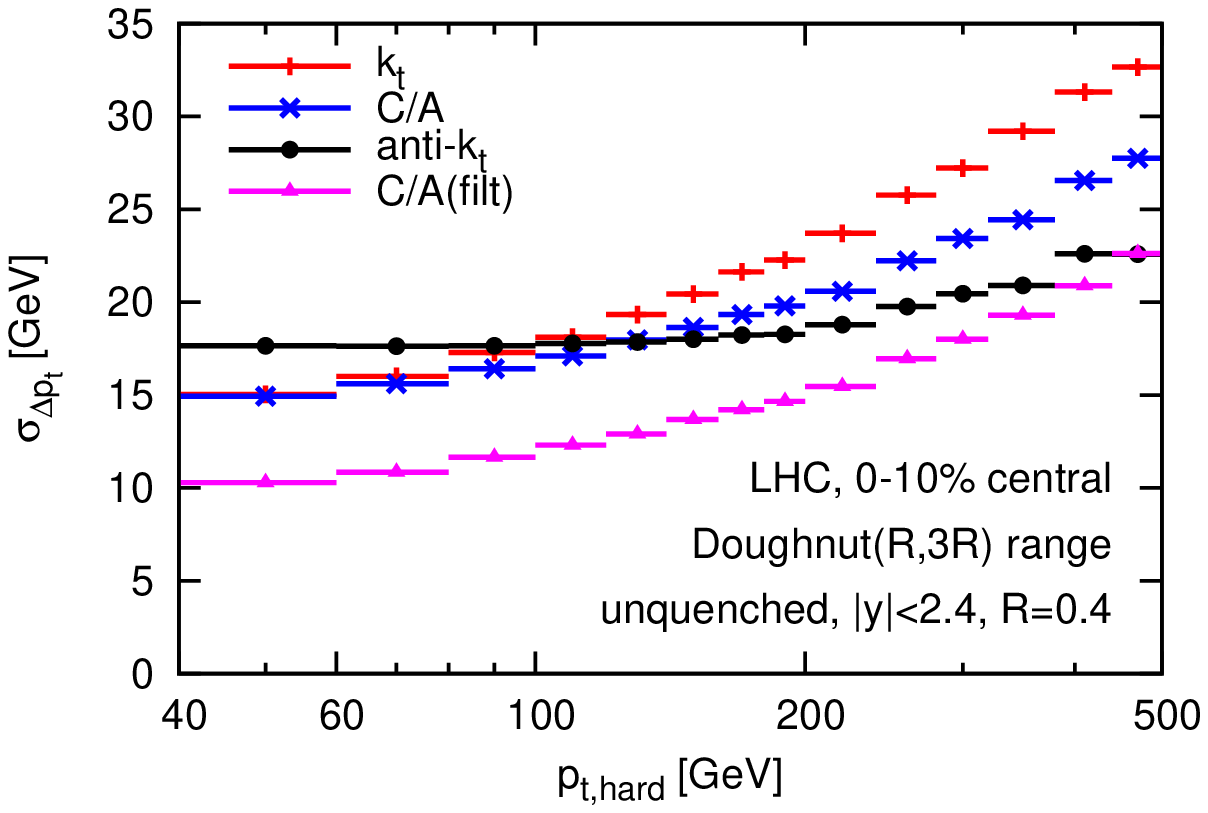}
}
\caption{\label{fig:dispersion} Dispersion $\sigma_{\Delta p_t}$. Left: RHIC, right: LHC.}
\end{figure}

Our results for the $\Delta p_t$ dispersion, $\sigma_{\Delta p_t}$,
are shown in fig.~\ref{fig:dispersion}, again using the doughnut
range. 
(Our conclusions are essentially independent of the particular choice
of range.)

We first discuss the case of RHIC kinematics. For $k_t$ and
anti-$k_t$, the observed dispersions are similar to the result of
$6.8\GeV$ quoted by STAR~\cite{star} (though the number from STAR
includes detector resolution effects, so that the true physical
$\sigma_{\Delta p_t}$ may actually be somewhat lower).
Of note, the advantage enjoyed
by anti-$k_t$ in terms of smallest $\avg{\Delta p_t}$ does not hold at
the level of the dispersion: C/A and $k_t$ tend to behave slightly
better at small transverse momentum. The algorithm which performs best
in terms of dispersion over all the $p_t$ range is now C/A with
filtering, for which the result is smaller than that of the other algorithms
by a factor of about $1/\sqrt{2}$.
This reduction factor can be explained because the dispersion
$\sigma_{\Delta p_t}$ is expected to be proportional to the
square-root of the jet area:
the C/A(filt) algorithm with $R_{\rm filt}=R/2$ and $n_{\rm filt}=2$
produces jets with an area of, roughly, half that obtained with
C/A;
hence the observed reduction of $\sigma_{\Delta p_t}$.

In the LHC setup, the conclusions are quite similar at the lowest
transverse momenta shown.
As $p_t$ increases, the dispersion of the anti-$k_t$ algorithm grows
slowly, while that of the others grows more rapidly, so that at the
highest $p_t$'s shown, the $k_t$ and C/A algorithms have noticeably
larger dispersions than anti-$k_t$, and C/A(filt) becomes similar to
anti-$k_t$.
The growth of the dispersions can be attributed to an increase of the
back-reaction dispersion.
The latter is dominated by rare occurrences, where a large fraction of
the jet's $p_t$ is gained or lost to back-reaction, hence the
noticeable $p_t$ dependence (\cnf
appendix~\ref{app:dispersion-contributions}).
An additional effect, especially for the $k_t$ algorithm, might come
from the anomalous dimension of the jet areas, \ie the growth with
$p_t$ of the average jet area.

Note that the dispersion has some limited dependence on the choice of
ghost area --- for example, reducing it from 0.01 to 0.0025 lowers
the dispersions by about $0.2-0.4\GeV$ at RHIC. This is discussed
further in appendix~\ref{app:ghost-area-choice}.

%
%
%
%
%

%
%
%
%
%
%
%
%
%

%
\subsection{Centrality dependence}\label{sec:centrality}

\begin{figure}
\centerline{\includegraphics[angle=270,width=0.6\textwidth]{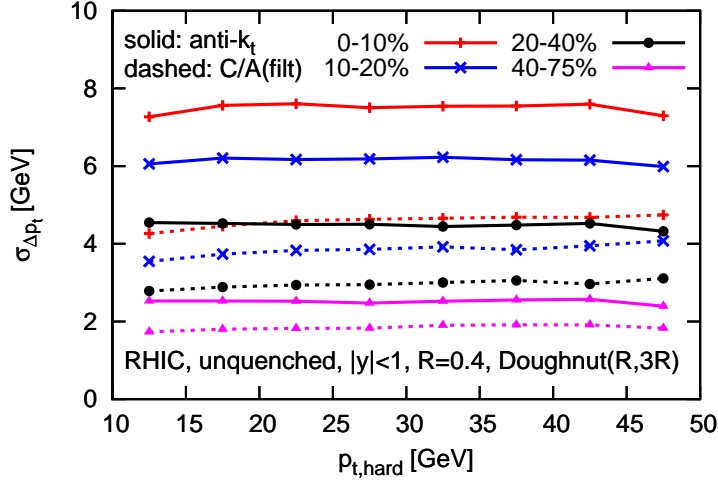}}
\caption{\label{fig:dispersion_centrality}
  $p_t$ dependence dependence of the $\Delta p_t$ dispersion at RHIC,
  for different centrality classes.}
\end{figure}

So far, we have only considered central collisions. Since it is known
that non-central collisions give rise to elliptic flow \cite{flow_exp,
  flow_th}, one might worry that this leads to an extra source of
background fluctuations and/or non-uniformities, potentially spoiling
the subtraction picture discussed so far.
One can study this on azimuthally averaged jet samples (as we have
been doing so far) or as a function of the azimuthal angle, $\Delta
\phi$, between the jet and the reaction plane.
As above, we use HYDJET~v1.6, whose underlying HYDRO component
includes a simulation of elliptic flow~\cite{Lokhtin:2003ru}.

We have generated heavy-ion background events for RHIC in four
different centrality bins: 0-10\% (as above), 10-20\%, 20-40\% and
40-75\%, with $v_2$ values respectively of $1.7\%$, $3.3\%$, $5.0\%$
and $5.3\%$.%
\footnote{$v_2$ was determined as the average of $\cos2\phi$ over all
  all particles with $|\eta| < 1$ (excluding the additional hard $pp$
  event).  }

We first examine azimuthally averaged results,
repeating the studies of the previous sections for each of the
centrality bins.
%
%
%
%
%
%
%
%
We find that the results for the average shift, $\avg{\Delta p_t}$,
are largely independent of centrality, as expected if the elliptic
flow effects disappear when averaged over $\phi$.
The  results for the dispersion are
shown in fig.~\ref{fig:dispersion_centrality}. 
We observe that the dispersion decreases with increasing
non-centrality. Even though one might expect adverse effects from
elliptic flow, the heavy-ion background decreases rapidly when one
moves from central to peripheral collisions, and this directly
translates into a decrease of $\sigma_{\Delta p_t}$.

The first conclusion from this centrality-dependence study is
therefore that the subtraction methods presented in this paper appear
to be applicable also for azimuthally averaged observables in
non-central collisions.

\begin{figure}
  \centerline{
    \includegraphics[angle=270,width=0.5\linewidth]{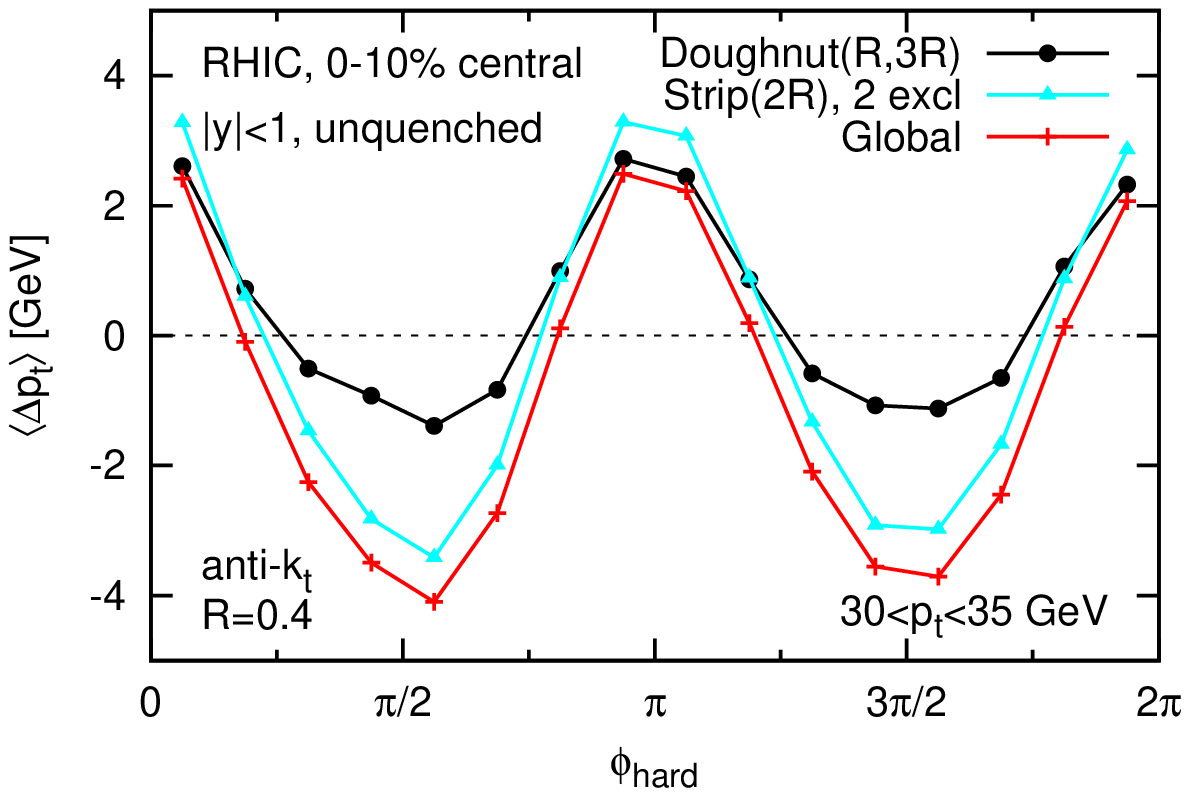}
    \includegraphics[angle=270,width=0.5\linewidth]{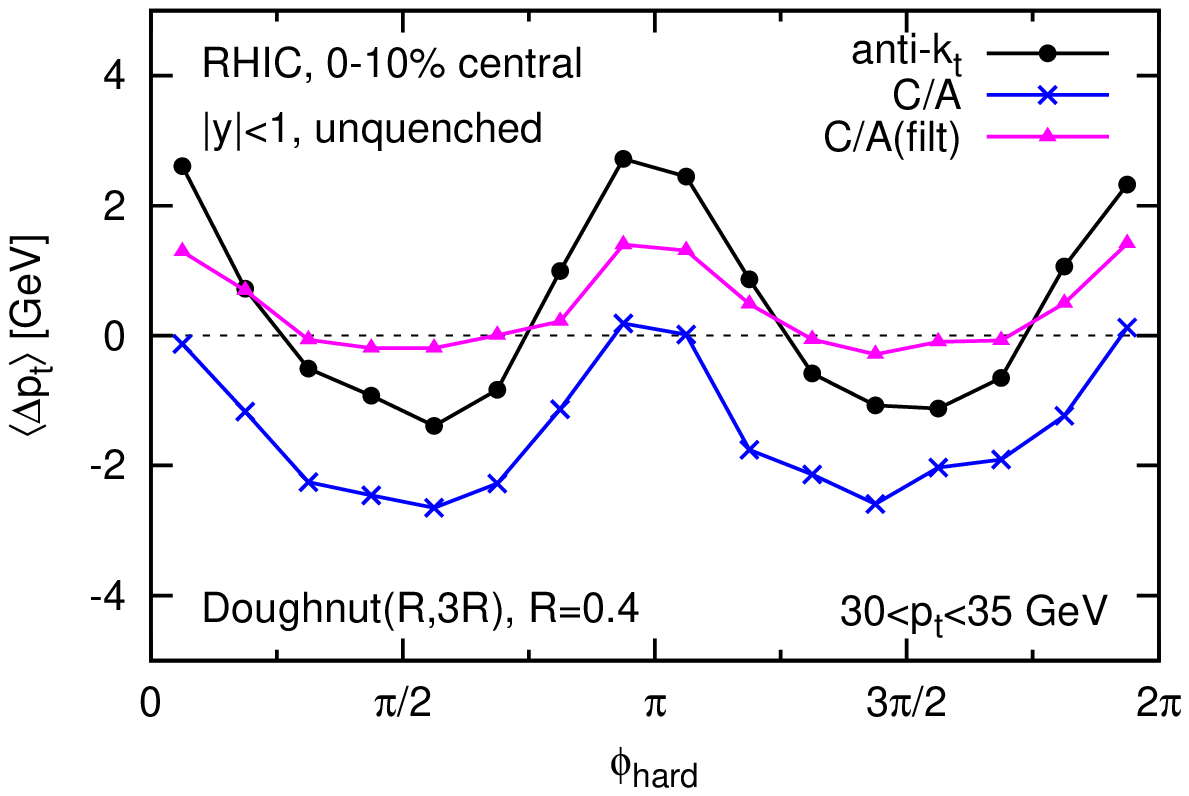}
  }
  \caption{\label{fig:ptshift-v-deltaphi}
    $\phi$ dependence of the $\Delta p_t$ shift at RHIC,
    for the $0-10\%$ centrality bin. Left: for three different ranges
    for the anti-$k_t$ algorithm;
    right: for three different jet algorithms for the doughnut($R,3R$) range.
    The absolute size of the $\phi$ dependence is similar for
    centralities up to $40\%$ and then decreases beyond.  }
\end{figure}

We next consider results as a function of $\Delta \phi$, which is
relevant if one wishes to examine the correlation between jet
quenching and the reaction plane.
An issue in real experimental studies is the determination of the
reaction plane, and the extent to which it is affected by the
presence of hard jets.
In the HYDJET simulations, this problem does not arise because the
reaction plane always corresponds to $\phi=0$.
Figure~\ref{fig:ptshift-v-deltaphi} (left) shows the average $\Delta p_t$ as
a function of $\Delta \phi$ for the anti-$k_t$ algorithm and several
different background-estimation ranges, for the $0-10\%$
centrality bin for RHIC.
The strip range shows significant $\Delta \phi$ dependence, which is
because a determination of $\rho$ averaged over all $\phi$ cannot
possibly account for the local $\phi$-dependence induced by the
elliptic flow.
Other ranges, such as the doughnut range, instead cover a more limited
region in $\phi$. They should therefore be able to provide information
on the $\phi$-dependence of the background.%
\footnote{At the expense of being more strongly affect by jet-medium
  interactions that could manifest themselves as broad enhancement of
  the energy flow in the vicinity of the jet.}
However, since their extent in $\phi$ tends to be significantly larger
than that of the jet, and $\rho$ varies relevantly over that extent,
some residual $\phi$ dependence remains in $\avg{\Delta p_t}$ after
subtraction.
The right-hand plot of figure~\ref{fig:ptshift-v-deltaphi} shows that
the effect is reduced with filtering, as is to be expected since its
initial background contamination is smaller.
The conclusion from this part of the study is that residual
$\phi$-dependent offsets may need to be corrected for explicitly in
any studies of jets and their correlations with the reaction plane.
The investigation of extensions to our background subtraction
procedure to address this issue will be the subject of future work.

%
\subsection{Quenching effects}\label{sec:quenching}

The last issue we wish to investigate is how  the phenomenon of  jet
quenching (\ie medium effects on parton fragmentation) may affect the
picture developed so far. 
The precise nature of jet quenching beyond its basic analytic properties
(see \eg \cite{bdmps}) is certainly hard to estimate in detail, especially
at the LHC, where experimental data from, say, flow or particle spectra
measurements are not yet available for the tuning of the Monte Carlo
simulations. Additionally, the implementation of Monte-Carlo generators
that incorporate the  analytic features of jet quenching models  is
currently a very active field \cite{qpythia,jewel,martini}. We may
therefore  expect a more robust and complete picture of jet quenching in the
near future, together with the awaited first \PbPb collisions at the LHC.

In this section, we examine the robustness of our HI background
subtraction in the presence of (simulated) jet quenching.\footnote{
  Our focus here is therefore not the study of quenching itself, but
  merely how it may affect our subtraction procedure.  }
For this purpose we have used two available models which allow one to
simulate quenched hard jets, PYQUEN \cite{pyquen}, which is used by
HYDJET~v1.6, and QPYTHIA \cite{qpythia}. PYQUEN has been run with the
parameters listed in footnote \ref{foot:pyquen} for the LHC, and with 
$\texttt{T0}=0.5\GeV$, $\texttt{tau0}=0.4$~fm and $\texttt{nf}=2$ for
RHIC\footnote{The parameters for RHIC are taken from \cite{pyquen}. 
The difference with the default parameters does not 
appear to be large for the purpose of our investigations and, in any 
case, a systematic study of quenching  effects is not among the goals 
of this paper.}.
For QPYTHIA we have
tested two options for the values of the transport coefficient and the
medium length ($\hat q = 3$~GeV$^2$/fm, $L = 5$~fm and $\hat q =
1$~GeV$^2$/fm, $L = 6$~fm), with similar results.
No serious attempt has been made to tune the two codes with each other
or with the experimental data, beyond what is already suggested by the
code defaults: in the absence of strong experimental constraints
on the details of the quenching effects, this allows us to verify the
robustness of our results for a range of conditions.

As in sections \ref{sec:choice_range} and \ref{sec:choice_alg}, we
have embedded the hard PYQUEN or QPYTHIA events in a HYDJET~v1.6
background and tested the effectiveness of the background subtraction
for different choices of algorithm.\footnote{The effect of the choice
  of range remains as in section \ref{sec:choice_range} for the
  unquenched case. We will therefore keep employing the doughnut
  range.  }
We shall restrict our attention to the anti-$k_t$ and C/A(filt)
algorithms, as they appear to be the optimal choices from our analysis
so far.

%
%
%
%
%
%
%
%
%
%
%
%
%
%
%
%
%
%
%
%
%

\begin{figure}
\centerline{
\includegraphics[angle=270,width=0.5\textwidth]{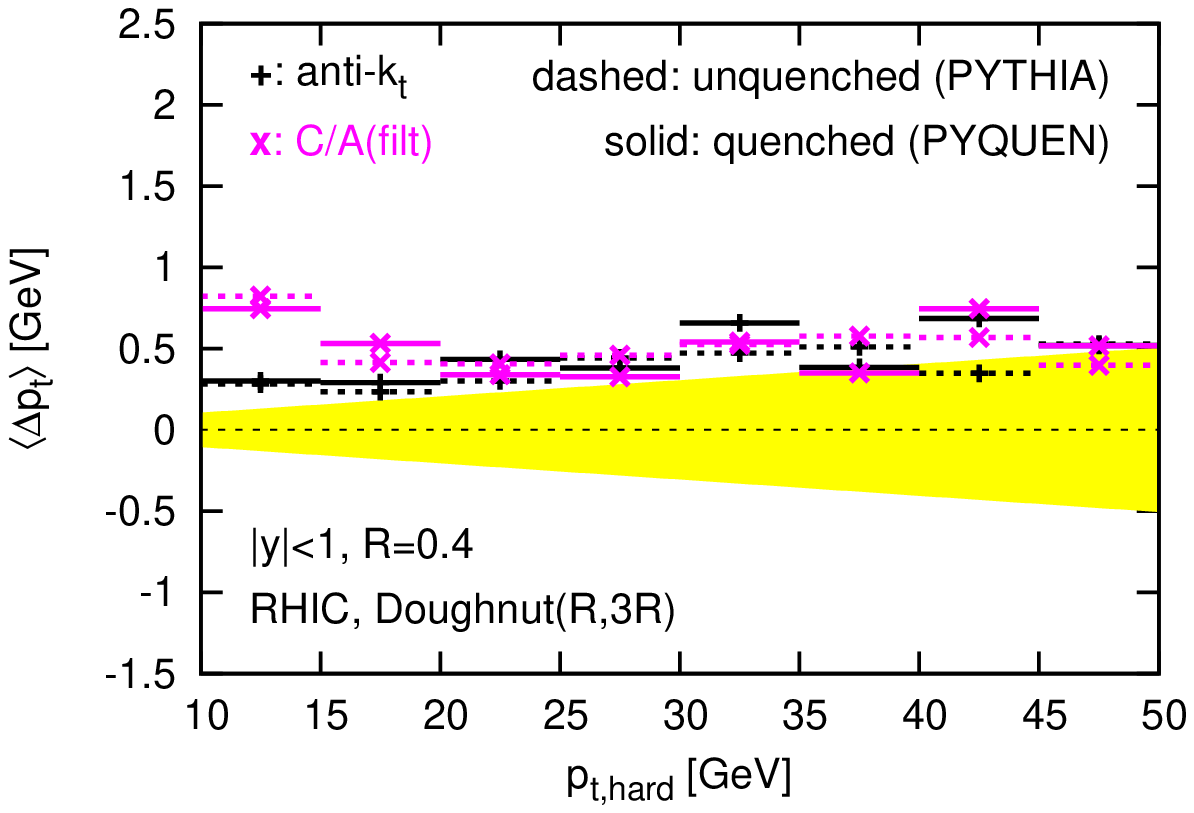}
\includegraphics[angle=270,width=0.5\textwidth]{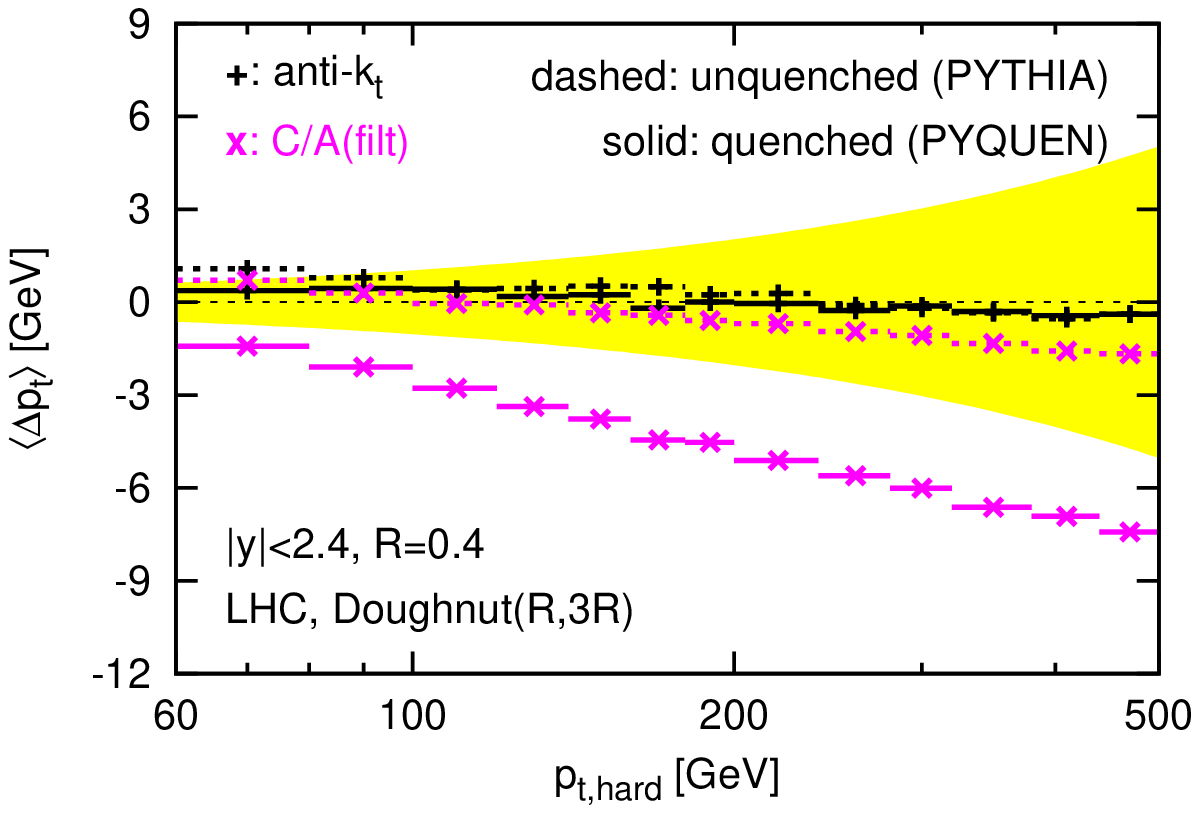}
}
\caption{\label{fig:ptshift_quenched}
  Average $p_t$ shift for background-subtracted jets with the
  anti-$k_t$ and C/A(filt) jet algorithms. The dashed lines correspond
  to unquenched hard jets (PYTHIA) and the solid ones to quenched hard
  jets (PYQUEN). Results are shown for RHIC kinematics on the left
  plot and for the LHC on the right one.}
\end{figure}

We have found that the jet-matching efficiencies are still high, with
essentially no changes at RHIC, and at LHC a doubling of the (small)
inefficiencies that we saw in fig.~\ref{fig:efficiency}(right).
The dispersions $\sigma_{\Delta p_t}$ are also not significantly
affected within our sample of jet-quenching simulations.
We therefore concentrate
on the $\avg{\Delta p_t}$ offset, which is plotted in
fig.~\ref{fig:ptshift_quenched} for PYQUEN.
The results are shown for both RHIC and the LHC.
In the case of RHIC, and for the whole $p_t$ range up to about 50 GeV,
quenching can be seen not to significantly affect the subtraction
offset $\langle \Delta p_t\rangle$ (within the usual uncertainty
related to the choice of range, which was shown in
fig.~\ref{fig:ptshift_ranges}). In the LHC case, instead, while the
shift obtained using the anti-$k_t$ algorithm is largely similar to
the unquenched case,
the C/A(filt) algorithm performance can be seen to deteriorate slightly when
quenching is turned on, all the more so at very large transverse momentum.
%
%
%
%
%
%
%
%
The $k_t$ and the C/A algorithms are not shown for clarity, but they
share the behaviour of C/A(filt).  This
deterioration of the quality of the subtraction can be traced back to
an increased back-reaction compared to the unquenched jets.
Anti-$k_t$ jets do not suffer from this effect as a consequence of the
usual rigidity of this algorithm.
In the case of C/A(filt), one should nevertheless emphasise that an
error of (at most) 10 GeV on the reconstruction of a 500 GeV jet is
still only a 2\% effect. This is modest, both relative to the likely
experimental precision and to the expected effect of quenching on the
overall jet $p_t$, predicted by PYQUEN to be at the level of $10\%$ at
this $p_t$.

Though for brevity we have not explicitly shown them, the results
with QPYTHIA are very similar.

Before closing this section, we reiterate that we have only
investigated simple models for quenching and that our results are 
meant just to give a first estimate of the effects that one might have to
deal with in the case of quenched jets. The expected future
availability of new ``quenched'' Monte Carlo programs, together with
specific measurements in the early days of heavy-ion collisions at the
LHC, will certainly allow one to address this question more
extensively.

%
%

\subsection{Relative importance of average shift and dispersion}
\label{eq:shift-dispersion-on-spectrum}

To close this section, we examine the relative importance of the
average shift and its
dispersion, taking the illustrative example of their impact on the
inclusive jet cross-section as a function of $p_t$. 
We start from a simple parametrisation of the inclusive-jet $p_t$
spectrum and see how its reconstruction is affected by the average
shift and dispersion that remain for $\Delta p_t$ after subtraction.

Let us assume that the true $p_t$ spectrum decays exponentially \ie
\begin{equation} 
  \label{eq:hard-spectrum}
  \frac{d\sigma^{pp}}{dp_t} = \sigma_0\lambda\,e^{-\lambda p_t}.
\end{equation}
While this expression doesn't have the $1/p_t^n$ form that one expects
to see, it is far easier to handle analytically, and not too poor an
approximation to observed spectra over quite a broad range of $p_t$.
After embedding the hard events in a heavy-ion background and applying
subtraction, the resulting spectrum will be the convolution of
eq.~(\ref{eq:hard-spectrum}) with the $\Delta p_t$
distribution. Assuming that the latter is a Gaussian of average
$\avg{\Delta p_t}$ and dispersion $\sigma_{\Delta p_t}$ (\cnf
fig.~\ref{fig:deltapt-dist}), one obtains a reconstructed spectrum
\begin{equation}
  \label{eq:smeared-hard-spectrum}
  \frac{d\sigma^{AA,\rm sub}}{dp_t} = 
  \exp\left(\lambda\avg{\Delta p_t}+\frac{\lambda^2\sigma_{\Delta p_t}^2}{2}\right)
  \, \frac{d\sigma^{pp}}{dp_t}.
\end{equation}
Thus the average shift gives a bias by a multiplicative factor
$\exp(\lambda\avg{\Delta p_t})$ and the dispersion by a factor
$\exp(\lambda^2\sigma_{\Delta p_t}^2/2)$. 
The convolution works in such a way that for a given reconstructed
$p_t$, the most likely original true transverse momentum is:
\begin{equation}
  \label{eq:most-likely}
  \text{most likely}\; \pthard \simeq 
   \ptfullsub - \avg{\Delta p_t} - \lambda\sigma_{\Delta p_t}^2\,,
\end{equation}
where we have neglected the small impact of subtraction on the $pp$
jets.

To illustrate these effects quantitatively, let us first take the
example of RHIC, where between 10 and 60 GeV, the cross-section is
well approximated by eq.~(\ref{eq:hard-spectrum}) with $\lambda=0.3$
GeV$^{-1}$. Both the anti-$k_t$ and C/A(filt) have $\avg{\Delta
  p_t}\simeq 0$, leaving only the dispersion effect. In the case of
the anti-$k_t$ (respectively C/A(filt)) algorithm, we see from
fig.~\ref{fig:dispersion} that $\sigma_{\Delta p_t}\simeq 7.5 \GeV$
($4.8\GeV$), which gives a multiplicative factor of about 12 (3).
For a given reconstructed $p_t$, the most likely true $p_t$ is about
$17\GeV$ ($7\GeV$) smaller.
In comparison, for C/A ($k_t$), with $\avg{\Delta p_t}\simeq -1.5\GeV$
($-3.5\GeV$) (fig.~\ref{fig:ptshift}) and $\sigma_{\Delta p_t}\simeq
6.5 \GeV$ (similar for $k_t$) there is a partial compensation between
factors of $0.64$ ($0.35$) and $6.7$ coming respectively from the
shift and dispersion, yielding an overall factor of about $4$ ($2.3$),
while the most likely true $p_t$ is about $14\GeV$ ($12\GeV$) smaller
than the reconstructed $p_t$.

At the LHC ($\sqrt{s_{NN}} = 5.5\TeV$), eq.~(\ref{eq:hard-spectrum})
is a less accurate approximation. Nevertheless, for $p_t \sim
100-150\GeV$, it is not too unreasonable to take $\lambda=0.05\GeV^{-1}$
and examine the consequences. For anti-$k_t$ (respectively C/A(filt)),
we have $\avg{\Delta p_t} \simeq 0$ (also for C/A(filt)) and
$\sigma_{\Delta p_t} \simeq 18\GeV$ ($13\GeV$), giving a
multiplicative factor of $1.5$ ($1.2$), i.e.\ far smaller corrections
than at RHIC.
For a given reconstructed $p_t$, the most likely true $p_t$ is about
$16\GeV$ ($8\GeV$) smaller, rather similar to the values we found at
RHIC (though smaller in relative terms, since the $p_t$'s are higher),
with the increase in $\sigma$ being compensated by the decrease in
$\lambda$.

In the LHC case, it is also worth commenting on the results for the
$k_t$ algorithm, since this is what was used in
ref.~\cite{subtraction}: we have $\avg{\Delta p_t} \simeq -8\GeV$ and
$\sigma_{\Delta p_t} \simeq 18\GeV$, giving a multiplicative factor of
$1.05$, which is consistent with the near perfect agreement that was
seen there between the $pp$ and subtracted $AA$ spectra.
That agreement does not however imply perfect reconstruction, since
the most likely $\pthard$ is about $7\GeV$ lower than $\ptfullsub$.

Though the above numbers give an idea of the relative difficulties of
using different algorithms at RHIC and LHC, 
experimentally what matters most will be the systematic errors on the
correction factors (for example due to poorly understood non-Gaussian
tails of the $\Delta p_t$ distribution). Note also that a compensation
between shift and dispersion factors, as happens for example with the
C/A algorithm, is unlikely to reduce the overall systematic errors.

%
%
%
%

\section{The issue of fakes}\label{sec:fakes}

While the goal of this paper is not to discuss the issue of
``fake-jets'' in detail, it is a question that has been the subject of
substantial debate recently (see for example
\cite{Lai:2008zp,JacobsPragueTalk}). Here, therefore, we wish to
devote a few words to it and discuss how it relates to our
background-subtraction results so far.

In a picture in which the soft background and the hard jets are
independent of each other, one way of thinking about a fake jet is
that it is a reconstructed jet (with significant $p_t$) that is due
not to the presence of an actual hard jet, but rather due to an
upwards fluctuation of the soft background.
The difficulty with this definition is that there is no
uniquely-defined separation between ``hard'' jets and soft background.
This can be illustrated with the example of how HYDJET
simulates RHIC collisions: one event typically consists of a soft
HYDRO background supplemented with $\sim 60$ $pp$ collisions, each simulated
with a minimum $p_t$ cut of $2.6\GeV$ on the $2\to2$ scattering.
To some approximation, the properties of the full heavy-ion events
remain relatively unchanged if one modifies the number of $pp$
collisions and corresponding $p_t$ cut and also retunes the soft
background. 
The fact that this changes the number of hard jets provides one
illustration of the issue that the soft/hard separation is
ill-defined.
Additionally, while there are $\sim 60$ semi-hard $pp$ collisions
($\sim 120$ mostly central semi-hard jets\footnote{Specifically,
  keeping in mind the HYDJET simulation, one can cluster each $pp$ event
  separately to obtain a long list of $pp$ jets from all the separate
  hard events.}) in an event, there is only space within (say) the
acceptance of RHIC for $\order{40}$ jets. Thus there is essentially no
region in an event which does not have a semi-hard jet.
From this point of view, every reconstructed jet corresponds to a
\mbox{(semi-)}hard $pp$ jet and there are no fake jets at all.

\subsection{Inclusive analyses}
\label{sec:fakes:incl}

For inclusive analyses, such as a measurement of the inclusive jet
spectrum, this last point is particularly relevant, because every jet
in the event contributes to the measurement. 
Then, the issue of fakes can be viewed as one of unfolding. 
In that respect it becomes instructive, for a given bin of the 
reconstructed heavy-ion $p_t$, to ask what the corresponding matched
$pp$ jet transverse momenta were.
Specifically, we define a quantity $O(\ptfullsub, p_{t}^{pp})$,
the distribution of the $pp$ ``origin'', $p_{t}^{pp}$, of a heavy-ion
jet with subtracted transverse momentum $\ptfullsub$.
If the origin $O(\ptfullsub, p_{t}^{pp})$ is dominated by a
region of $p_{t}^{pp}$ of the same order as $\ptfullsub$, then
that tells us that the jets being reconstructed are truly hard.
If, on the other hand, it is dominated by $p_{t}^{pp}$ near zero, then
that is a sign that apparently hard heavy-ion jets are mostly due to
upwards fluctuations of the background superimposed on low-$p_t$ $pp$
jets, making the unfolding more delicate.

\begin{figure}
  \includegraphics[angle=-90,width=\textwidth]{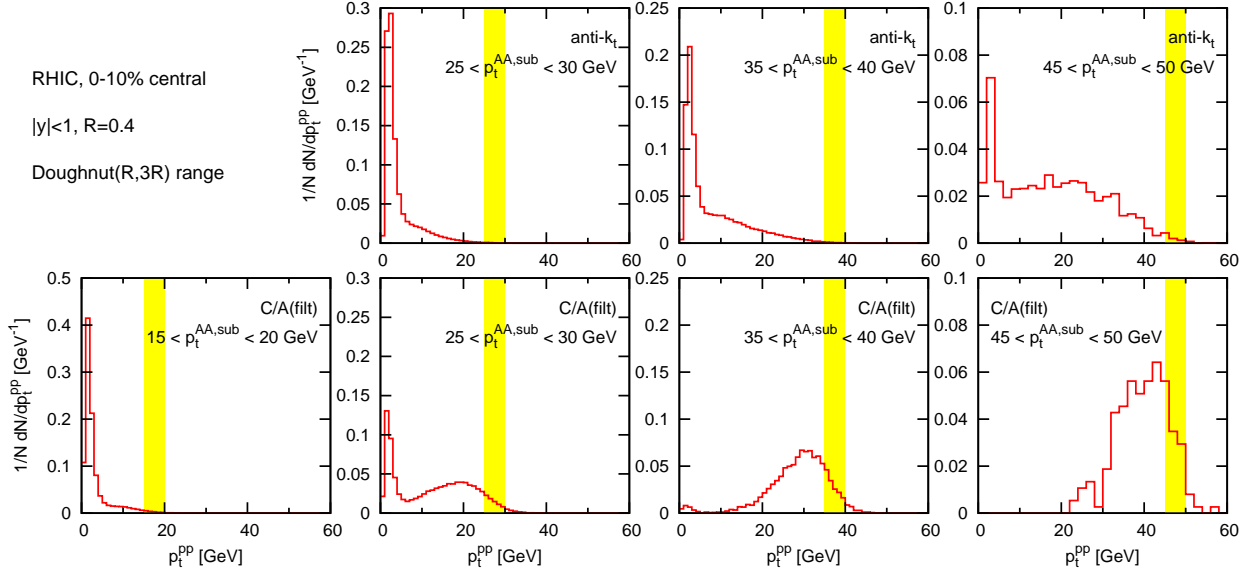}
  \caption{The $p_t$ distribution of the $pp$ jet corresponding to a
    given bin of reconstructed heavy-ion jet $\ptfullsub$ at RHIC, i.e.\ 
    $O(\ptfullsub, p_{t}^{pp})$ as a function of $p_{t}^{pp}$
    for a given bin of $\ptfullsub$.
    The upper row is for the anti-$k_t$ algorithm, while the lower row
    is for C/A(filt). 
    Each column corresponds to a different $\ptfullsub$ bin, as
    indicated by the vertical band in each plot. 
    Cases in which the histogram is broad or peaked near $0$ are
    indicative of the need for special care in the unfolding
    procedure. 
    These plots were generated using approximately 90 million events.
    Each plot has been normalised to the number of events in the
    corresponding $\ptfullsub$ bin.
  }
  \label{fig:origin}
\end{figure}

\begin{figure}
  \includegraphics[angle=-90,width=\textwidth]{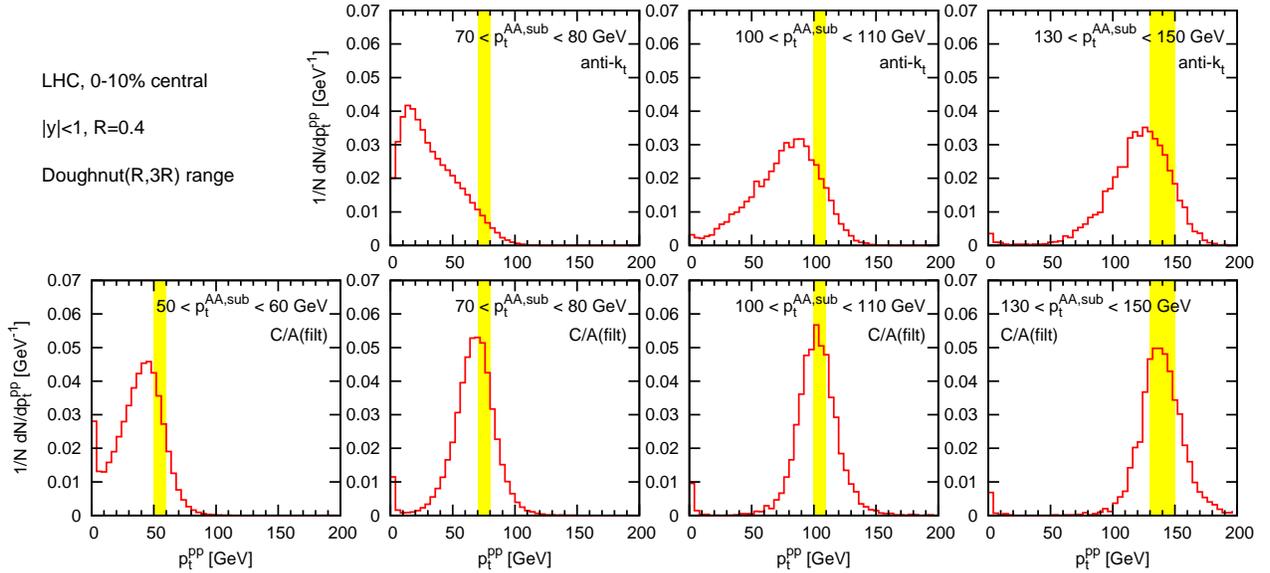}
  \caption{Same as fig.~\ref{fig:origin} for LHC kinematics (PbPb,
    $\sqrt{s} = 5.5\TeV$), generated with approximately 16 million
    events.
    Note the use of a smaller rapidity range here, $|y|<1$, compared to
    the earlier LHC plots. 
  }
  \label{fig:origin_lhc}
\end{figure}

In figure~\ref{fig:origin} we show the origins of heavy-ion jets as
determined in our HYDJET simulations.\footnote{%
A point to be aware of is that multiple $pp$ jets can match a single
heavy-ion jet, i.e.\ have at least half their $p_t$ contained in the
heavy-ion jet.
In evaluating $O(\ptfullsub, p_{t}^{pp})$ we take only the
highest-$p_t$ matched jet. If there is no matched jet (this occurs
only rarely) then we fill the bin at $p_{t}^{pp} = 0$.
Note also that since we are not explicitly embedding hard jets, all
$pp$ jets in the events have undergone HYDJET's quenching.
}
The upper row provides the origin plots for anti-$k_t$ jets at
RHIC. Each plot corresponds to one bin of $\ptfullsub$, and
shows $O(\ptfullsub, p_{t}^{pp})$ as a function of $p_{t}^{pp}$.
At moderate $\ptfullsub$, the $25-30\GeV$ bin, the origin is
dominated by low $\pthard$.
This is perhaps not surprising, given the result in
section~\ref{eq:shift-dispersion-on-spectrum} that the $\pthard$
origin is expected to be $\sim 17\GeV$ lower than $\ptfullsub$
for anti-$k_t$ jets --- additionally, that result assumed an
exponential spectrum for the inclusive jet distribution, whereas the
distribution rises substantially faster towards low $\pthard$.
As $\ptfullsub$ increases one sees that the contribution of
high $\pthard$ jets increases, in a manner not too inconsistent with
the expected $\sim 17\GeV$ shift, though the $\pthard$ distribution
remains rather broad and a peak persists at small $\pthard$.
These plots suggest that an inclusive jet distribution measurement
with the anti-$k_t$ algorithm at RHIC is not completely
trivial 
since, up to rather large $\ptfullsub$, one is still sensitive to the
jet distribution at small values of $\pthard$ where the separation
between ``hard'' jets and the soft medium is less clear.
Nevertheless, two points should be kept in mind: firstly, the upper
row of fig.~\ref{fig:origin} shows that different $\ptfullsub$
have complementary sensitivities to different parts of the $\pthard$
spectrum. Thus it should still be possible to ``unfold'' the
$\ptfullsub$ distribution to obtain information about the
$\pthard$, unfolding being in any case a standard part of the
experimental correction procedure.
Secondly STAR quotes~\cite{star} a $10\%$ smaller value for
$\sigma_{\Delta p_t}$ than the $7.5\GeV$ that we find in HYDJET. Such
a reduction can make it noticeably easier to perform the unfolding.

The impact of a reduction in $\sigma_{\Delta p_t}$ is illustrated in
the lower row of fig.~\ref{fig:origin}, which shows the result for
C/A(filt).
Here, even the $25-30\GeV$ bin for $\ptfullsub$ shows a
moderate-$p_t$ peak in the distribution of $\pthard$, and in the
$35-40\GeV$ bin the low-$p_t$ ``fake'' peak has disappeared almost
entirely. 
Furthermore, the $\pthard$ peak is centred about $7\GeV$ lower than
the centre of the $\ptfullsub$ bin, remarkably consistent with
the calculations of section~\ref{eq:shift-dispersion-on-spectrum}.
Overall, therefore, unfolding with C/A(filt) will be easier than
with anti-$k_t$.

Corresponding plots for the LHC are shown in
fig.~\ref{fig:origin_lhc}.
While C/A(filt)'s lower dispersion still gives it an advantage over
anti-$k_t$, for $\smash{\ptfullsub} \gtrsim 80\GeV$, anti-$k_t$ does
now reach a domain where the original $pp$ jets are themselves always
hard.

Procedures to reject fake jets have been proposed,
in~\cite{Grau:2008ed,Lai:2009ai}. They are based on a cut on
(collinear unsafe) jet shape properties and it is thus unclear how
they will be affected by quenching and in particular whether the
expected benefit of cutting the low-$\pthard$ peak in
figs.~\ref{fig:origin} and \ref{fig:origin_lhc} outweighs the
disadvantage of potentially introducing extra sources of systematic
uncertainty at moderate $p_t$.

One final comment is that experimental unfolding should provide enough
information to produce origin plots like those shown here. As part of
the broader discussion about fakes it would probably be instructive
for such plots to be shown together with the inclusive-jet results.
%

\subsection{Exclusive analyses}
\label{sec:fakes:excl}

An example of an exclusive analysis might be a dijet study, in which
one selects the two hardest jets in the event, with transverse momenta
$p_{t1}$ and $p_{t2}$, and plots the distribution of $\frac12 H_{T,2}
\equiv \frac12(p_{t1} + p_{t2})$.
Here one can define ``fakes'' as corresponding to cases where one or
other of the jets fails to match to one of the two hardest among all
the jets from the individual $pp$ events.
This definition is 
insensitive to the soft/hard boundary in a
simulation such as HYDJET, because it naturally picks out hard $pp$ 
jets that are far above that boundary. And, by concentrating on just
two jets, it also evades the problem of high occupancy from the large
multiplicity of semi-hard $pp$ collisions.
This simplification of the definition of fakes is common to many
exclusive analyses, because they tend to share the feature of
identifying just one or two hard reference jets.

The specific case of the exclusive dijet analysis has the added
advantage that it is amenable to a data-driven estimation of
fakes. 
One divides the events into two groups, those for which the two
hardest jets are on the same side (in azimuth) of the event and those
in which they are on opposite sides (a related analysis was presented
by STAR in ref.~\cite{BrunaPrague}).
For events in which one of the two jets is ``fake,'' the two jets are
just as likely to be on the same side as on the opposite side. This is
not the case for non-fake jets, given that the two hardest ``true''
jets nearly always come from the same $pp$ event and so have to be on
opposite sides.\footnote{%
  At RHIC energies, above $p_t\sim 10-15\GeV$, it is nearly always the
  case that the two hardest jets come from the same $pp$
  event. At the LHC, this happens above $20-30\GeV$.}
Thus by counting the number of same-side versus opposite-side dijets
in a given $\frac12 H_{T,2}$ bin, one immediately has an estimate of
the fake rate.\footnote{Note that for plain $pp$ events, if one has
  only limited rapidity acceptance then the same-side/opposite-side
  separation is not infrared safe, because of events in which only one
  hard jet is within the acceptance and the other ``jet'' is given by
  a soft gluon emission.
  Thus to examine the same-side/opposite-side separation in plain $pp$
  events with limited acceptance, one would need to impose a $p_t$ cut
  on the second jet, say $p_{t,2} > \frac12 p_{t,1}$.  }

\begin{figure}[t]
  \centering
  \includegraphics[height=0.5\textwidth,angle=-90]{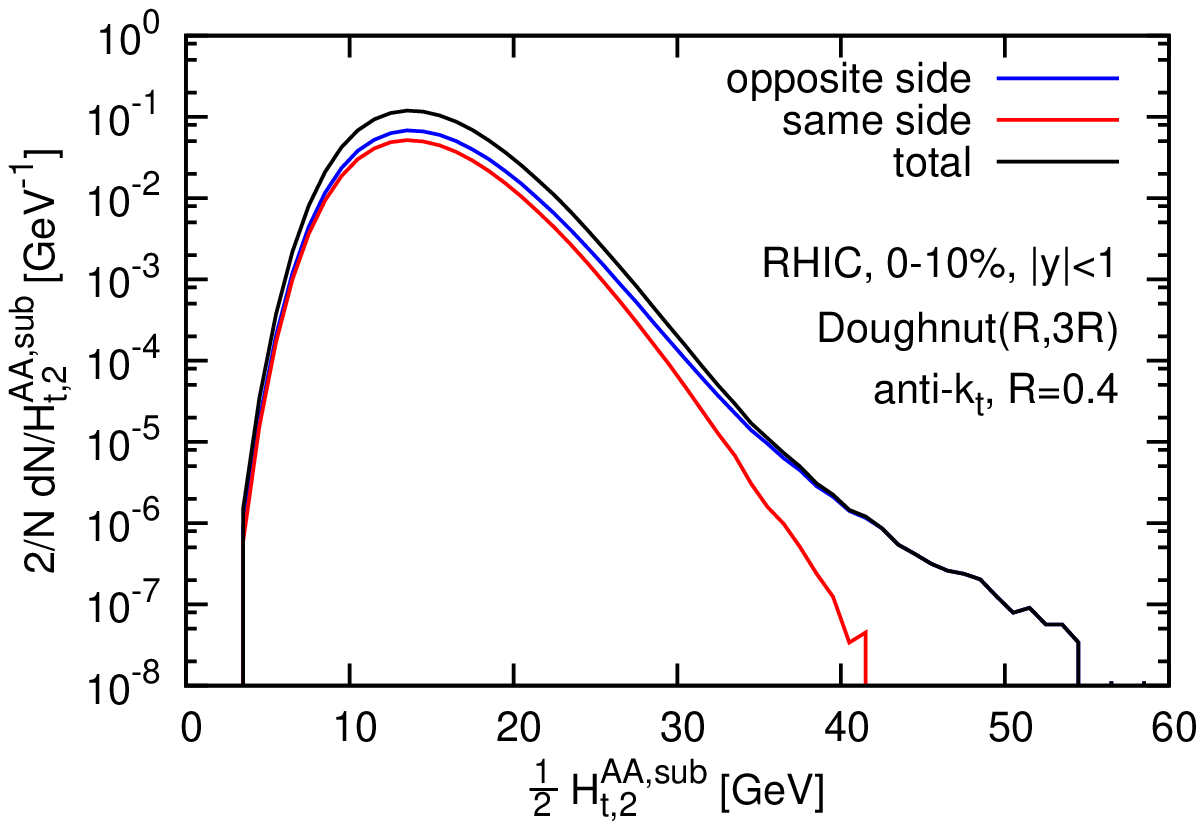}%
  \includegraphics[height=0.5\textwidth,angle=-90]{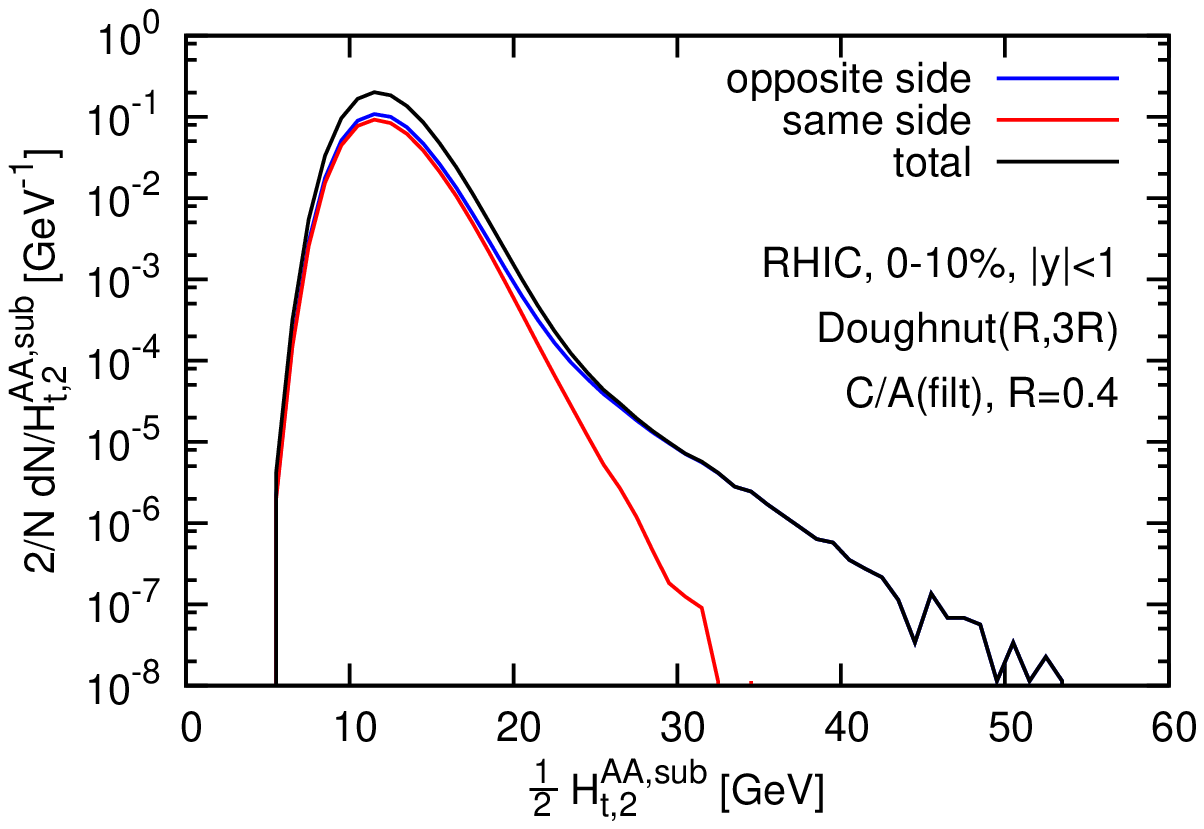}%
  \caption{The distribution of $\frac12 H_{T,2}$ obtained from the two
    hardest full, subtracted heavy-ion jets in each event at RHIC, as
    obtained from simulations with HYDJET 1.6. The left-hand plot is
    for anti-$k_t$ and the right-hand plot for C/A(filt).
    The same-side curves give an approximate measure of (half of) the
    contribution of ``fake'' jets to the dijet spectrum.  }
  \label{fig:dijets}
  \centering
  \includegraphics[height=0.5\textwidth,angle=-90]{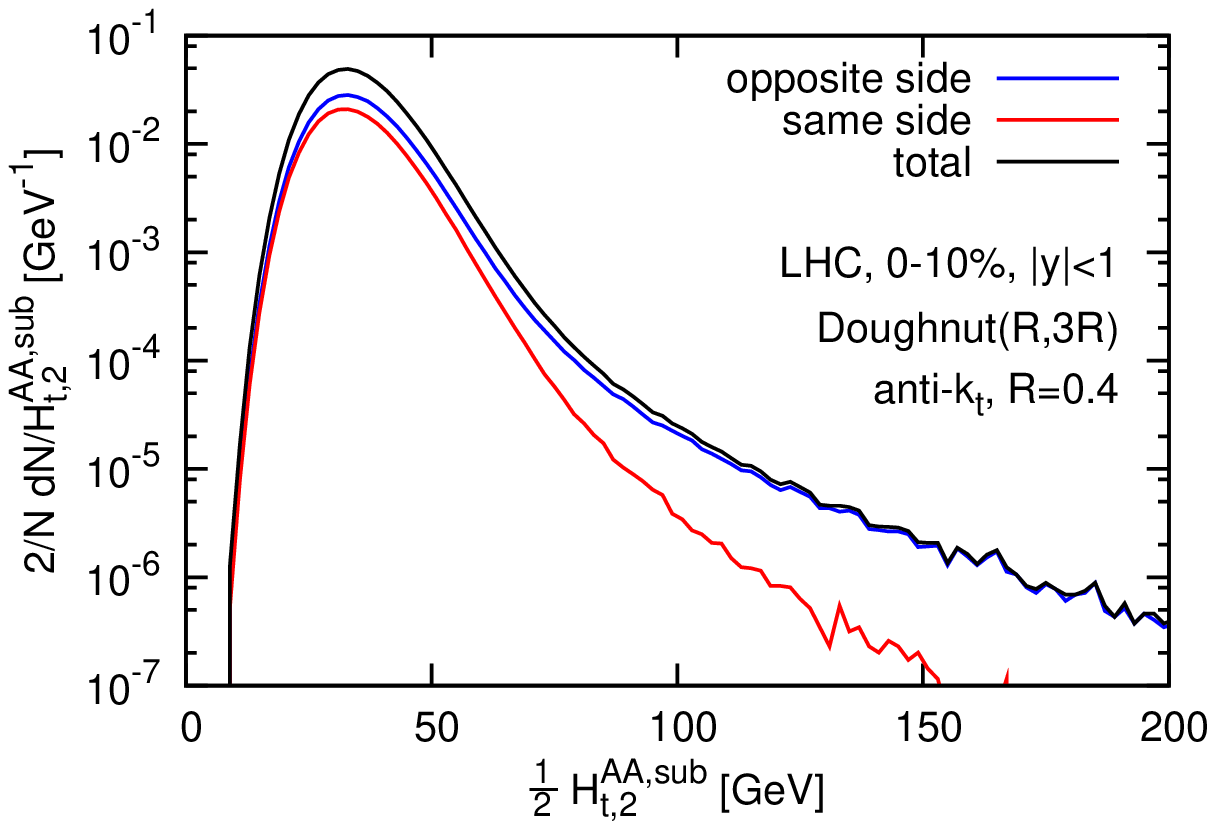}%
  \includegraphics[height=0.5\textwidth,angle=-90]{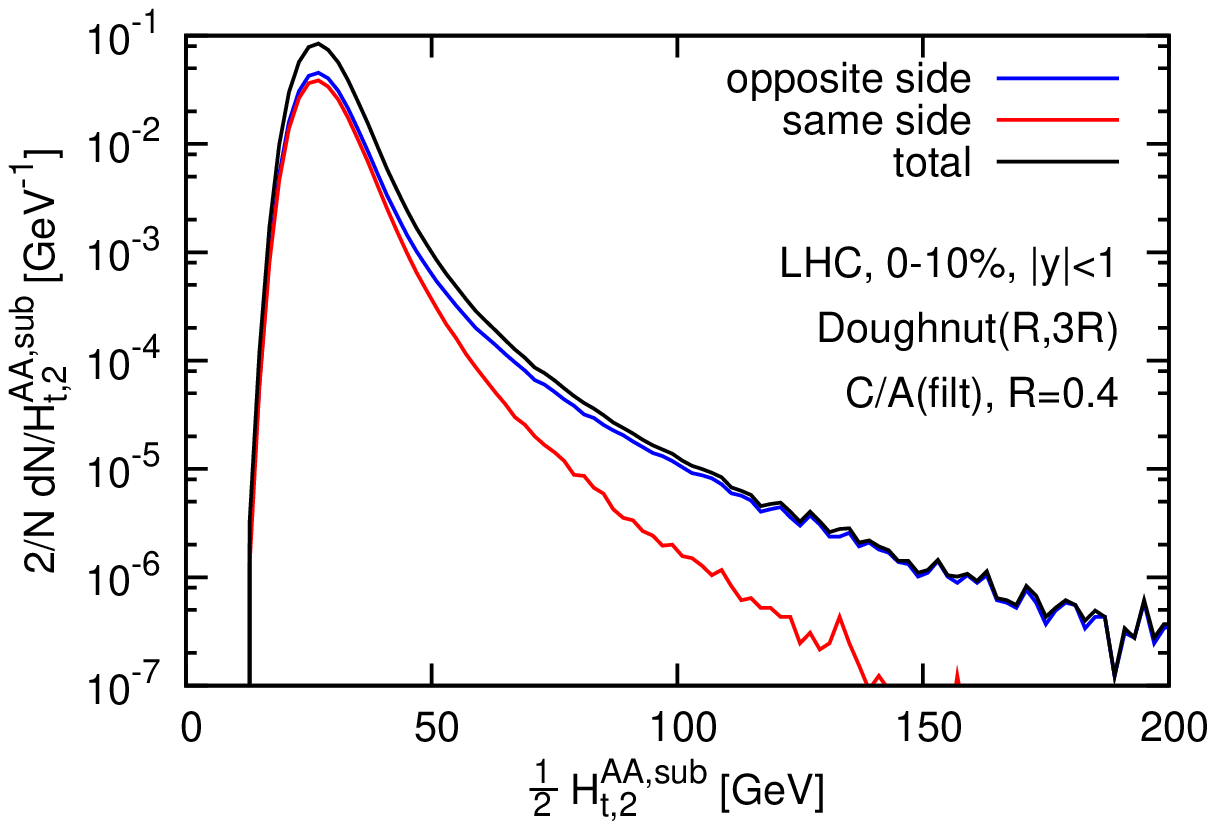}%
  \caption{Same as fig.~\ref{fig:dijets} for LHC kinematics.}
  \label{fig:dijets_lhc}
\end{figure}

This is illustrated in fig.~\ref{fig:dijets}, which shows the
distribution at RHIC of the full, subtracted $\frac12 H_{T,2}$ result,
together with its separation into opposite-side and same-side
components.
One sees that in the peak region, the opposite and same-side
distributions are very similar, indicating a predominantly ``fake''
origin for at least one of two hardest jets (they are not quite
identical, because there is less phase-space on the same side for a
second jet than there is on the away side).
However above a certain full, reconstructed $\frac12 H_{T,2}$ value,
about $30\GeV$ for anti-$k_t$ and $20\GeV$ for C/A(filt) the
same-side distribution starts to fall far more rapidly than the
opposite-side one, indicating that the measurement is now dominated by
``true'' pairs of jets. 

The LHC results, fig.~\ref{fig:dijets_lhc}, are qualitatively similar,
with the same-side spectrum starting to fall off more steeply than the
opposite-side one around $70\GeV$ for the anti-$k_t$ algorithm and $50\GeV$ for
C/A(filt).

One can also examine origin plots for $H_{T,2}$, in analogy with
the Monte Carlo analysis of section~\ref{sec:fakes:incl}. For
brevity, we refrain from showing them here, and restrict ourselves to
the comment that in the region of $H_{T,2}$ where the result is
dominated by opposite-side pairs, the origin plots are consistent with
a purely hard origin for the dijets.

%
\section{Conclusions}\label{sec:ccl}

In this article we have presented the results of a systematic study of
heavy-ion jet reconstruction with jet-area based background
subtraction, building on the brief initial proposal of
\cite{subtraction}.

The questions we have examined include those of the choice of range
for estimating the background, the choice of algorithm for the jet
finding and the robustness of the reconstruction with respect to
quenching effects and collision centrality. 

We have found that there is little difference between various ranges,
as long as they are chosen to be localised to the vicinity of the jet
of interest and of sufficient size (at least 4 units of area for jets
with $R=0.4$).
In comparing different algorithms we examined the systematic offset
and the dispersion in the reconstructed jet $p_t$.
The offset can be brought close to zero by using the anti-$k_t$
algorithm, while the $k_t$ algorithm has the largest offset; the
Cambridge/Aachen (C/A) algorithm with filtering also gives a small offset,
however this seems to have been due to a fortuitous cancellation
between two only partially related effects.
The dispersion is comparable for anti-$k_t$, C/A and $k_t$, but
significantly smaller for C/A(filt) (except at high transverse momenta
for LHC), as a consequence of its smaller jet area.
Among the different algorithms, anti-$k_t$ is the most robust with
respect to quenching effects, and C/A(filt) seems reasonably robust at
RHIC, though a little less so at the LHC.
The precise numerical results for offset and dispersion can depend a
little on the details of the simulation and the analysis, however the
general pattern remains.

Overall our results indicate that the area-based subtraction method
seems well suited for jet reconstruction in heavy-ion collisions.
Two jet-algorithm choices were found to perform particularly well:
anti-$k_t$, which has small offsets but larger fluctuations, and C/A
with filtering, for which the offsets may be harder to control, but
for which the fluctuations are significantly reduced, with consequent
advantages for the unfolding of experimentally measured jet
spectra.
Ultimately, we suspect that carrying out parallel analyses with these
two choices may help maximise the reliability of jet results in HI
collisions.

\section*{Acknowledgements}

We wish to thank N.~Armesto, H.~Caines, D.~d'Enterria, P.~Jacobs and
C.~Salgado for numerous helpful comments on the manuscript.
We are grateful to them and to B.~Cole, L.~Cunqueiro, Y-S.~Lai,
M.~Ploskon, J.~Puthschke, T.~Renk, C.~Roland, S.~Salur, U.~Wiedemann
and many others, for useful and stimulating conversations on the topic
of jets in heavy ion collisions over the past couple of years.
We also thank J.~Andersen for helpful suggestions to obtain some of
the computing resources used for this paper.
GS would like to acknowledge his affiliation to the Brookhaven
National Laboratory while a large part of this work was performed and
JR would like to acknowledge his affiliation to the LPTHE in Paris
during its initial stages.
This work has been supported by the French Agence Nationale de la
Recherche, under grants ANR-05-JCJC-0046-01 and ANR-09-BLAN-0060 and
the U.S. Department of Energy under contract DE-AC02-98CH10886.

\appendix

\section{Estimate of the minimal size of a range}
\label{sec:minimal-range}

\subsection{Fluctuations in extracted $\rho$}\label{app:minrange-fluct}

In section \ref{sec:subtraction}, we gave an estimate of the minimum
size of a range one should require for determining $\rho$, given a
requirement that fluctuations in the determination of $\rho$ should be
moderate.
We give the details of the computation in this appendix.

We start from the fact that the error made on the estimation of the
background density $\rho$ will translate into an increase of the
dispersion $\sigma_{\Delta p_t}$: on one hand, the dominant
contribution to $\sigma_{\Delta p_t}$ comes from the intra-event
fluctuations of the background \ie is of order $\sigma \sqrt{A_{\rm
    jet}}$ with $A_{\rm jet}$ the jet area; on the other hand, the
dispersion $S_{\Delta \rho}$ of the misestimation of $\rho$ leads to
an additional dispersion on the reconstructed jet $p_t$ of $S_{\Delta
  \rho} A_{\rm jet}$. Adding these two sources of dispersion in
quadrature and using the result from section~3.4 of \cite{css}, $S_{\Delta \rho}
\simeq \sqrt{\pi/(2 A_{\cal R})} \sigma$ with $A_{\cal R}$ the area of the
range under consideration, we get
\begin{equation}
\sigma_{\Delta p_t} \simeq \sigma \left(A_{\rm jet} 
  + \frac{\pi A_{\rm jet}^2}{2 A_{\cal R}}\right)^{\frac12}.\label{eq:1}
\end{equation}
If we ask \eg that the contribution to the total dispersion coming
from the misestimation of the background be no more than a fraction
$\epsilon$ of the total $\sigma_{\Delta p_t}$, then we obtain the
requirement
\begin{equation}
  \label{eq:AR-min}
  A_{\cal R} \gtrsim  A_{\min} \simeq \frac{\pi}{4} \frac{A_{\rm jet}}{\epsilon}\,.
\end{equation}
For anti-$k_t$ jets of radius $R$, with $A_{\rm jet} \simeq \pi R^2$,
this translates to
\begin{equation}
  \label{eq:AR-min-in-terms-of-R}
  A_{\cal R} \gtrsim \frac{\pi^2}{4} \frac{R^2}{\epsilon} \simeq 25 R^2\,,
\end{equation}
where the numerical result has been given for $\epsilon = 0.1$. For
$R=0.4$, it becomes $A_{\cal R} \gtrsim 4$.

We can also cast this result in terms of the number of jets that must
be present in ${\cal R}$. Assuming that the jets used to estimate
$\rho$ have a mean area of $0.55 \pi R_\rho^2$,\footnote{This is the
  typical area one would obtain using a (strongly recommended) jet
  definition like the $k_t$ or C/A algorithms for the background
  estimation~\cite{areas}.} we find a minimal number of jets\,,
\begin{equation}
  \label{eq:njets-min}
  n_{\min} \simeq \frac{A_{\min}}{0.55 \pi R_{\rho}^2} \simeq
  \frac{1.4}{\epsilon}\frac{R^2}{R_\rho^2}\,,
\end{equation}
where, as before, we have taken $A_{\jet}\simeq \pi R^2$.
Taking the numbers quoted above and $R_\rho = 0.5$, as used in the
main body of the article, this gives $n_{\min}\simeq 9$.

\subsection{Hard-jet bias in extracted $\rho$}\label{app:minrange-bias}

From section~3.3 of \cite{css}, we know that the presence of hard jets
and initial-state radiation leads to a bias in the extraction of
$\rho$ of
\begin{equation}
  \label{eq:rho-bias}
  \avg{\Delta \rho} \simeq \sigma R_\rho \sqrt{\frac{\pi c_J}{2}}
  \frac{\avg{n_h}}{A_{\cal R}}\,
\end{equation}
where $c_J\simeq 2$ is a numerical constant and $\avg{n_h}$ is the
average number of ``hard'' jets (those above the scale of the
background fluctuations, including initial-state radiation);
$\avg{n_h}$ is given by
\begin{equation}
  \label{eq:nhard}
  \frac{\avg{n_h}}{A_{\cal R}} \simeq \frac{n_b}{A_{\cal R}} +
    \frac{C_i}{\pi^2}\frac{L}{b_0}\,,
    \qquad\quad
    L = \ln \frac{\as\left(\sqrt{c_J} \sigma R_\rho\right)}{\as(p_t)}\,,
\end{equation}
where $n_b$ is the number of ``Born'' partons from the underlying $2\to 2$
scattering that enter the region $\cal R$, while $b_0=(11 C_A -
2n_f)/(12 \pi)$ is the first coefficient of the QCD $\beta$-function.
In the context of \cite{css}, principally directed towards a study of
the UE in $pp$ collisions, $\sigma$ was rather small, causing the
second term of eq.~(\ref{eq:nhard}), associated with initial-state
radiation above a scale $\sim \sigma R_\rho$, to be comparable in
size to the first term.
In our case, $\sigma$ is significantly larger and this reduces
the impact of the second term sufficiently that we can ignore it.
We thus arrive at the result 
\begin{equation}
  \label{eq:rho-bias-result}
  \avg{\Delta \rho} \simeq \sigma R_\rho \sqrt{\frac{\pi c_J}{2}}
  \frac{n_b}{A_{\cal R}} \simeq 1.8 \,\sigma R_\rho
  \frac{n_b}{A_{\cal R}}\,.
\end{equation}

Note that the presence of hard jets and initial-state radiation also
affects the fluctuations in the misestimation of $\rho$ and this
should in principle have been included in the estimates of
appendix~\ref{app:minrange-fluct}. 
However, while the effect is not completely negligible, to within the
accuracy that is relevant for us (a few tens of percent in the
estimation of a minimal $A_{\cal R}$) it does not significantly alter
the picture outlined there.

\section{Subtraction bias due to filtering}\label{app:filtbias}

We have seen from fig.~\ref{fig:ptshift} in section
\ref{sec:choice_alg} that the subtraction differs when we use the C/A
algorithm with and without filtering. Since this difference is not due
to back-reaction (see fig.~\ref{fig:ptshift_br}), it has to be due to
the subtraction itself.

The difference comes from a bias introduced by the selection
of the two hardest subjets during filtering. The dominant contribution
comes when only one subjet, that we shall assume harder than all the
others, contains the hard radiation, all the other subjets being pure
background. In that case, the selection of the hardest of these
pure-background subjets as the second subjet to be kept tends to
pick positive fluctuations of the background. This in turn results in
a positive offset compared to pure C/A clustering, as observed in
section \ref{sec:choice_alg}.

To compute the effect analytically, let us thus assume that we have
one hard subjet and $n_{\rm bkg}$ pure-background subjets of area ${\cal A}_g =
0.55 \pi R_{\rm filt}^2$~\cite{areas}. After subtraction, the momentum of each of
the pure-background subjet can be approximated as having a Gaussian
distribution of average zero and dispersion $\sigma \sqrt{{\cal
    A}_g}$.
Assuming that the ``hard'' subjet's transverse momentum remains larger
than that of all the background jets, the 2 subjets that will be kept
by the filter are the hard subjet (subtracted) and the hardest of
all the subtracted background jets. The momentum distribution of the
latter is given by the maximum of the $n_{\rm bkg}$ Gaussian
distributions.\footnote{%
  Within \fastjet's filtering tools, when the subtracted transverse
  momentum of a subjet is negative, the subjet is assumed to be pure
  noise and so discarded.
  This means that the momentum distribution of the hardest subtracted
  background jet is really given by the distribution of the maximum of
  the $n$ Gaussian-distributed random numbers, but with the result
  replaced by zero if all of them are negative.
  In the calculations here we ignore this subtlety, since we will have
  $n=3$ and only $1/8^\mathrm{th}$ of the time are three Gaussian-distributed
  random numbers all negative.}
We are only interested here in computing the average bias introduced
by the filtering procedure, which is then given by
\[
\left\langle (\Delta p_t)_{\rm filt}\right\rangle
 \simeq \int \prod_{k=1}^{n_{\rm bkg}} \left(dp_{t,k}\,\frac{1}{\sqrt{2\pi{\cal
         A}_g}\sigma}e^{-\frac{p_{t,k}^2}{2{\cal A}_g
       \sigma^2}}\right)\:{\rm max}\left(p_{t,1},\dots,p_{t,n_{\rm bkg}}\right)
\]
For the typical case $R_{\rm filt}=R/2$ and $n_{\rm bkg}=3$, one finds
\begin{equation}
\left\langle (\Delta p_t)_{\rm filt}\right\rangle
  \simeq \frac{3\sqrt{{\cal A}_g}\sigma}{2 \sqrt{\pi}} 
  \simeq 0.56\,R\sigma.
\end{equation}
If we insert in that expression the typical values for the
fluctuations quoted in section \ref{sec:results} and $R=0.4$, we find
average biases of 2~GeV for RHIC and 4.5~GeV at the LHC, which are
in good agreement with the differences observed between C/A with and
without filtering in fig.~\ref{fig:ptshift}.

%

\section{Contributions to dispersion}

\subsection{Back reaction versus background fluctuations}
\label{app:dispersion-contributions}
\begin{figure}
  \centering
  \includegraphics[height=0.5\textwidth,angle=-90]{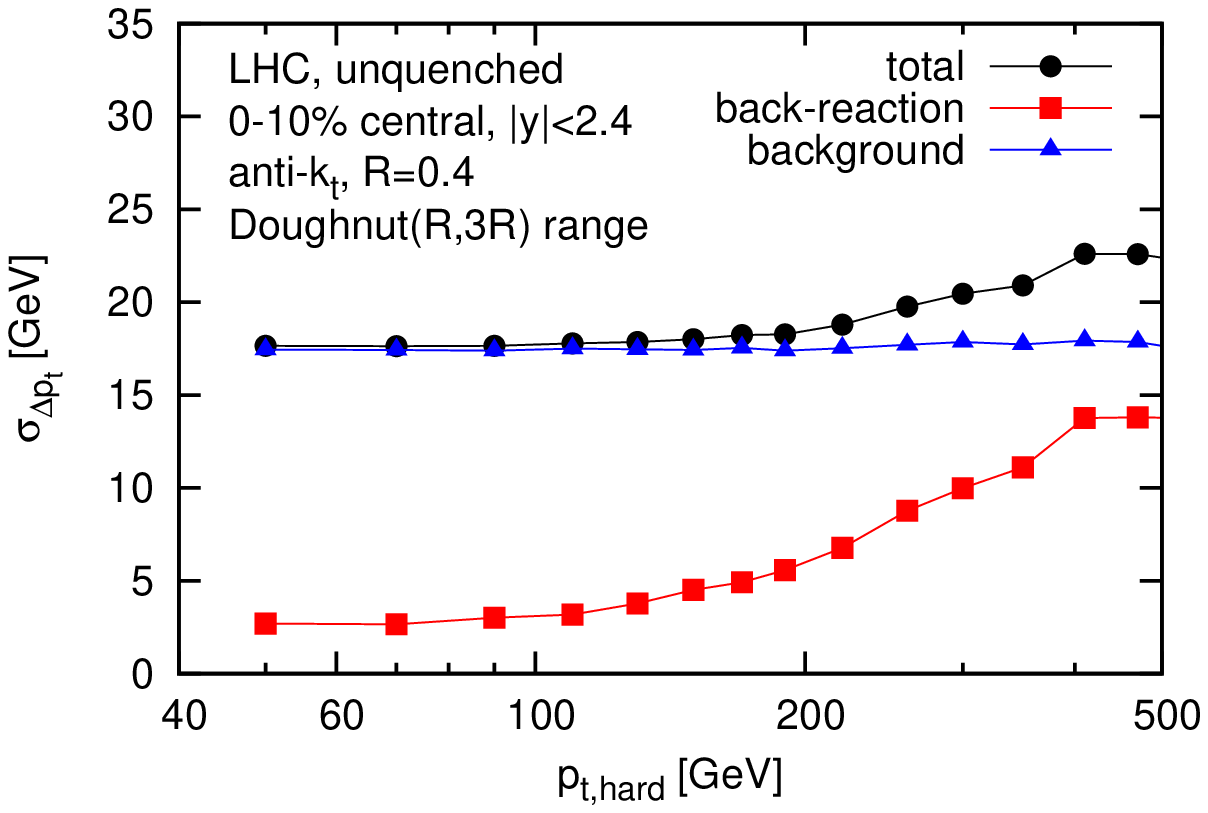}\hfill
  \includegraphics[height=0.5\textwidth,angle=-90]{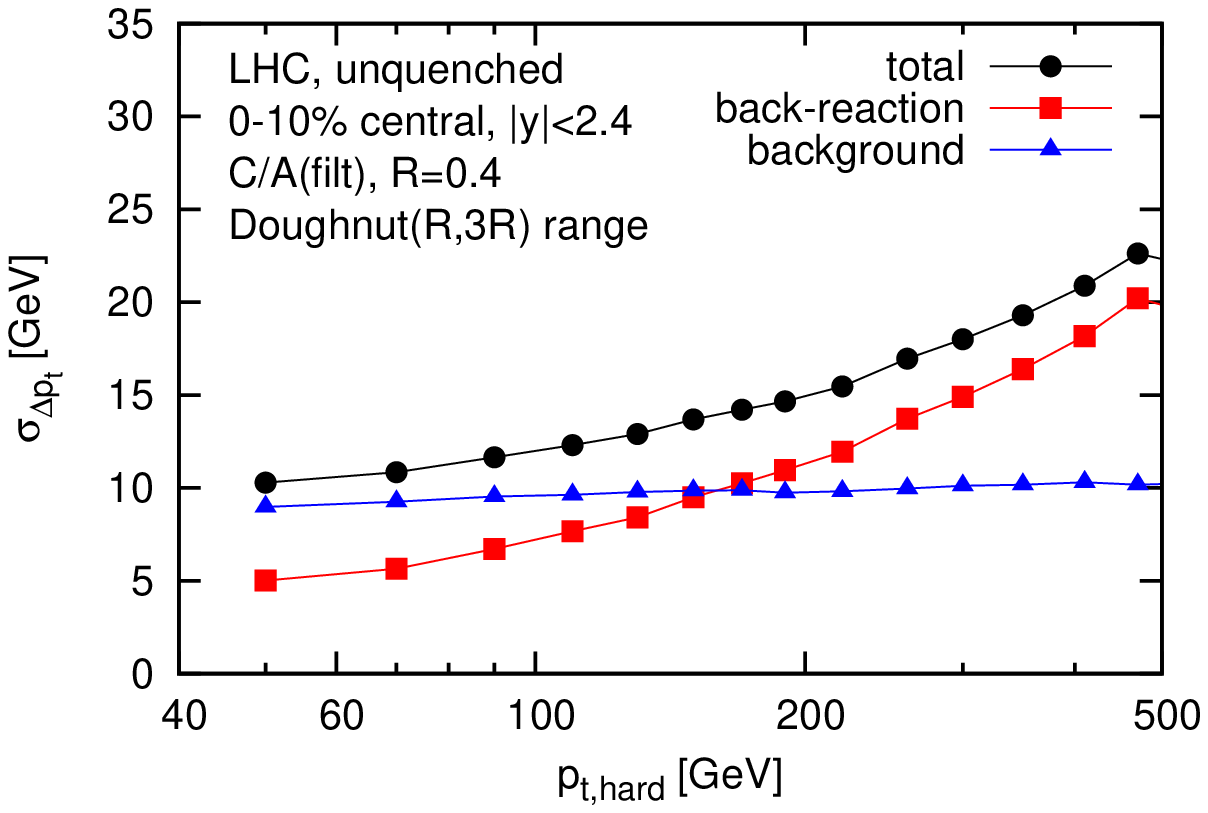}
  \caption{The decomposition of the dispersion into back-reaction and
    ``background'' components (including misestimation of
    $\rho$). The left-hand plot is for the anti-$k_t$ algorithm and
    the right-hand one for C/A(filt). Both correspond to LHC
    collisions at $\sqrt{s_{NN}} = 5.5\TeV$.}
  \label{fig:BR-v-full-dispersion}
\end{figure}

We stated, in section~\ref{sec:dispersion}, that the increase of the
dispersion at high $p_t$ seen in fig.~\ref{fig:dispersion} was mainly
due to back-reaction. This is made explicit in
fig.~\ref{fig:BR-v-full-dispersion}, which decomposes the dispersion
into its two components: that associated with the back-reaction,
$\sigma_{\Delta p_t}^\mathrm{BR}$ and that associated with background
fluctuations and misestimation of $\rho$ (defined as $[\sigma_{\Delta
  p_t}^2 - (\sigma_{\Delta p_t}^\mathrm{BR})^2]^{\frac12}$).
One sees that the background-fluctuation component is
essentially independent of $p_t$, while the back-reaction dispersion
has a noticeable $p_t$ dependence.
This is the case because the back-reaction dispersion is dominated by
rare events in which two similarly hard subjets are separated by a
distance close to $R$ (specifically by $R+\epsilon$ with $\epsilon \ll
1$). In such a configuration, the background's contribution to the
two subjets can affect whether they recombine and so lead to a large,
$\order{p_t}$, change to the jet's momentum.
In the limit of a uniform background, $\sigma/\rho \ll 1$, this can be
shown to occur with a probability of order $\as \rho R^2/p_t$. Thus
the contribution to the average shift $\avg{\Delta p_t}$ is
proportional to $\as \rho R^2$ (which in a full analysis is found to
be enhanced by a logarithm \cite{areas} for the $k_t$ and C/A
algorithms), while the contribution to $\avg{\Delta p_t^2}$ goes as
$\as \rho R^2 p_t$, and so leads to a dispersion that should grow
asymptotically as $\sqrt{p_t}$ (fig.~\ref{fig:BR-v-full-dispersion}
is, however, probably not yet in the asymptotic regime).

It is worth keeping in mind that even though rare but large
back-reaction dominates the overall dispersion, it will probably not
be the main contributor in distorting the reconstructed jet spectrum.
Such distortions come from upwards $\Delta p_t$ fluctuations, whereas
large back-reaction tends to be dominated by downwards fluctuations. The
reason is simple: in order to have an upwards fluctuation from
back-reaction, there must be extra $p_t$ near the jet in the original
$pp$ event. This implies the presence of a harder underlying $2\to 2$
scattering than would be deduced from the jet $p_t$, with a
corresponding significant price to pay in terms of more suppressed
matrix elements and PDFs.

Figure~\ref{fig:BR-v-full-dispersion} also shows that the
non-back-reaction component is nearly independent of $p_t$. This is
expected since the anomalous dimension of the jet area is zero for
anti-$k_t$ and small for C/A (with or without filtering), and in any
case leads to a weak scaling with $p_t$, as $\ln \ln p_t$.
Furthermore, there is roughly a factor of $1/\sqrt{2}$ between
anti-$k_t$ and C/A(filt), as expected based on a proportionality of
the dispersion to the square-root of the jet area.

\subsection{Quality of area determination}
\label{app:ghost-area-choice}

\begin{figure}
  \centering
  \includegraphics[height=0.6\textwidth,angle=-90]{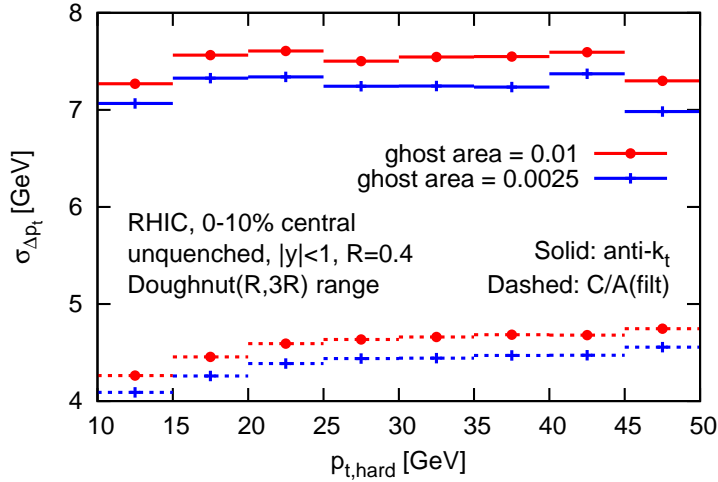}
  \caption{Dispersion, $\sigma_{\Delta p_t}$, for RHIC, as a function
    of the jet $p_t$, with two different choices for the ghost area,
    $0.01$ and $0.0025$. }
  \label{fig:ghost-area-impact}
\end{figure}

One further source of $\Delta p_t$ fluctuations can come from
imperfect estimation of the area of the jets. 
We recall that throughout this article we have used soft ghosts, each
with area of $0.01$, in order to establish the jet area. That implies
a corresponding finite resolution on the jet area 
and related poor estimation of the exact
edges of the jets, which can have an impact on the amount of
background that one subtracts from each jet, and, consequently, on
the final dispersion.
It is therefore interesting to see,
figure~\ref{fig:ghost-area-impact}, that the dispersion
$\sigma_{\Delta p_t}$ is reduced by about $0.2-0.4\GeV$ if one lowers
the ghost area to $0.0025$.

While this doesn't affect any of the conclusions of our paper, it does
suggest that for a full experimental analysis there are benefits to be
had from using a ghost area that is smaller than the default FastJet
setting of $0.01$.

\end{document}